\documentclass[12pt]{article}
\usepackage{myart}

\renewcommand{\theequation}{\arabic{section}.\arabic{equation}}
\normalbaselines
\input tcilatex

\begin{document}

\title{Deformation principle as foundation of physical geometry and its application
to space-time geometry.}
\author{Yuri A.Rylov}
\date{Institute for Problems in Mechanics, Russian Academy of Sciences,\\
101-1, Vernadskii Ave., Moscow, 119526, Russia.\\
e-mail: rylov@ipmnet.ru\\
Web site: {$http://rsfq1.physics.sunysb.edu/\symbol{126}rylov/yrylov.htm$}\\
or mirror Web site: {$http://gasdyn-ipm.ipmnet.ru/\symbol{126}%
rylov/yrylov.htm$}}
\maketitle

\begin{abstract}
Physical geometry studies mutual disposition of geometrical objects and
points in space, or space-time, which is described by the distance function $%
d$, or by the world function $\sigma =d^{2}/2$. One suggests a new general
method of the physical geometry construction. The proper Euclidean geometry
is described in terms of its world function $\sigma _{\mathrm{E}}$. Any
physical geometry $\mathcal{G}$ is obtained from the Euclidean geometry as a
result of replacement of the Euclidean world function $\sigma _{\mathrm{E}}$
by the world function $\sigma $ of $\mathcal{G}$. This method is very simple
and effective. It introduces a new geometric property: nondegeneracy of
geometry. Using this method, one can construct deterministic space-time
geometries with primordially stochastic motion of free particles and
geometrized particle mass. Such a space-time geometry defined properly (with
quantum constant as an attribute of geometry) allows one to explain quantum
effects as a result of the statistical description of the stochastic
particle motion (without a use of quantum principles).
\end{abstract}

\section{Introduction}

A geometry lies in the foundation of physics, and a true conception of
geometry is very important for the consequent development of physics. It is
common practice to think that all problems in foundations of geometry have
been solved many years ago. It is valid, but this concerns the geometry
considered to be a logical construction. Physicists are interested in the
geometry considered as a science on mutual disposition of geometrical
objects in the space or in the space-time. The two aspects of geometry are
quite different, and one can speak about two different geometries, using for
them two different terms. Geometry as a logical construction is a
homogeneous geometry, where all points have the same properties. Well known
mathematician Felix Klein \cite{K37} believed that only the homogeneous
geometry deserves to be called a geometry. It is his opinion that the
Riemannian geometry (in general, inhomogeneous geometry) should be qualified
as a Riemannian geography, or a Riemannian topography. In other words, Felix
Klein considered a geometry mainly as a logical construction. We shall refer
to such a geometry as the mathematical geometry.

The geometry considered to be a science on mutual disposition of geometric
objects will be referred to as a physical geometry, because the physicists
are interested mainly in this aspect of a geometry. The physical geometries
are inhomogeneous, in general, although they may be homogeneous also. On the
one hand, the proper Euclidean geometry is a physical geometry. On the other
hand, it is a logical construction, because it is homogeneous and can be
constructed of simple elements (points, straights, planes, etc.). All
elements of the Euclidean geometry have similar properties, which are
described by axioms. Similarity of geometrical elements allows one to
construct the mathematical (homogeneous) geometry by means of logical
reasonings. The proper Euclidean geometry was constructed many years ago by
Euclid. Consistency of this construction was investigated and proved in \cite
{H30}. Such a construction is very complicated even in the case of the
proper Euclidean geometry, because simple geometrical objects are used for
construction of the more complicated ones, and one cannot construct a
complicated geometrical object $\mathcal{O}$ without construction of the
more simple constituents of this object.

Note that constructing his geometry, Euclid did not use coordinates for
labeling of the space points. His description of the homogeneous geometry
was coordinateless. It means that the coordinates are not a necessary
attribute of the geometry. Coordinate system is a method of the geometry
description, which may or may not be used. Application of coordinates and of
other means of description poses the problem of separation of the geometry
properties from the properties of the means of the description. Usually the
separation of the geometry properties from the coordinate system properties
is carried out as follows. The geometry is described in all possible
coordinate systems. Transformations from one coordinate system to the
another one form a group of transformation. Invariants of this
transformation group are the same in all coordinate system, and hence, they
describe properties of the geometry in question.

At this point we are to make a very important remark. \textit{Any geometry
is a totality of all geometric objects }$\mathcal{O}$\textit{\ and of all
relations }$\mathcal{R}$\textit{\ between them}. Any geometric object $%
\mathcal{O}$ is a subset of points of some point set $\Omega $, where the
geometry is given. In the Riemannian geometry (and in other inhomogeneous
geometries) the set $\Omega $ is supposed to be a $n$-dimensional manifold,
whose points $P$ are labelled by $n$ coordinates $x=\left\{
x^{1},x^{2},...x^{n}\right\} $. This labelling (arithmetization of space) is
considered to be a necessary attribute of the Riemannian geometry. Most
geometers believe that the Riemannian geometry (and physical geometry), in
general cannot be constructed without introduction of the manifold. In other
words, they believe that the manifold is an attribute of the Riemannian
geometry (and of any continuous geometry, in general). This belief is
founded on the fact, that the Riemannian geometry is always constructed on
some manifold. But this belief is a delusion. The fact, that we always
construct the physical geometry on some manifold, does not mean that the
physical geometry cannot be constructed without a reference to a manifold,
or to a coordinate system. Of course, some labelling of the spatial points
(coordinate system) is convenient, but this labelling has no relations to
the construction of the geometry, and the physical geometry should be
constructed without a reference to coordinate system. Application of the
coordinate system imposes constraints on properties of the constructed
physical geometry. For instance, if we use a continuous coordinate system
(manifold) we can construct only continuous physical geometry. To construct
a discrete physical geometry, the geometry construction is not to contain a
reference to the coordinate system.

Here we present the method of the physical geometry construction, which does
not contain a reference to the coordinate system and other means of
description. It contains a reference only to the distance function $d$,
which is a real characteristic of physical geometry.

If a geometry is inhomogeneous, and the straights located in different
places have different properties, it is impossible to describe properties of
straights by means of axioms, because there are no such axioms for the whole
geometry. Mutual disposition of points in a physical (inhomogeneous)
geometry, which is given on the set $\Omega $ of points $P$, is described by
the distance function $d\left( P,Q\right) $%
\begin{equation}
d:\qquad \Omega \times \Omega \rightarrow \mathbf{R},\qquad d\left(
P,P\right) =0,\qquad \forall P\in \Omega  \label{b1.1}
\end{equation}
where $\mathbf{R}$ denotes the set of all real numbers. The distance
function $d$ is the main characteristic of the physical geometry. Besides,
the distance function $d$ is an \textit{unique characteristic} of any
physical geometry. \textit{The distance function }$d$\textit{\ determines
completely the physical geometry, and one does not need any additional
information for determination of the physical geometry.} This statement is
very important for construction of a physical geometry. It will be proved
below. Any physical geometry $\mathcal{G}$ is constructed on the basis of
the proper Euclidean geometry $\mathcal{G}_{\mathrm{E}}$ by means of a
deformation, i.e. by a replacement of the Euclidean distance function $d_{%
\mathrm{E}}$ by the distance function of the geometry in question. For
instance, constructing the Riemannian geometry, we replace the Euclidean
infinitesimal distance $dS_{\mathrm{E}}=\sqrt{g_{\mathrm{E}ik}dx^{i}dx^{k}}$
by the Riemannian one $dS=\sqrt{g_{ik}dx^{i}dx^{k}}$. There is no method of
the inhomogeneous physical geometry construction other, than the deformation
of the Euclidean geometry (or some other homogeneous geometry) which is
constructed as a mathematical geometry on the basis of its axiomatics and
logic. Unfortunately, conventional method of the Riemannian geometry
construction contains a reference to the coordinate system. But this
reference can be eliminated, provided that we use finite distances $d$
instead of infinitesimal distances $dS$.

For description of a physical geometry one uses the world function $\sigma $ 
\cite{S60}, which is connected with the distance function $d$ by means of
the relation $\sigma \left( P,Q\right) =\frac{1}{2}d^{2}\left( P,Q\right) $.
The world function $\sigma $ of the $\sigma $-space $V=\left\{ \sigma
,\Omega \right\} $ is defined by the relation 
\begin{equation}
\sigma :\qquad \Omega \times \Omega \rightarrow \mathbf{R},\qquad \sigma
\left( P,P\right) =0,\qquad \forall P\in \Omega  \label{b1.2}
\end{equation}
where $\mathbf{R}$ denotes the set of all real numbers. Application of the
world function is more convenient in the relation that the world function is
real, when the distance function $d$ is imaginary and does not satisfy
definition (\ref{b1.1}). It is important at the consideration of the
space-time geometry as a physical geometry.

In general, a physical geometry cannot be constructed as a logical building,
because any change of the world function should be accompanied by a change
of axiomatics. This is practically aerial, because the set of possible
physical geometries is continual. Does the world function contain full
information which is necessary for construction of the physical geometry? It
is a very important question. For instance, can one derive the space
dimension from the world function in the case of Euclidean geometry?
Slightly below we shall answer this question in the affirmative. Now we
formulate the method of the physical geometry construction.

Let us imagine that the proper Euclidean geometry $\mathcal{G}_{\mathrm{E}}$
can be described completely in terms and only in terms of the Euclidean
world function $\sigma _{\mathrm{E}}$. Such a description is called $\sigma $%
-immanent. It means that any geometrical object $\mathcal{O}_{\mathrm{E}}$
and any relation $\mathcal{R}_{\mathrm{E}}$ between geometrical objects in $%
\mathcal{G}_{\mathrm{E}}$ can be described in terms of $\sigma _{\mathrm{E}}$
in the form $\mathcal{O}_{\mathrm{E}}\left( \sigma _{\mathrm{E}}\right) $
and $\mathcal{R}_{\mathrm{E}}\left( \sigma _{\mathrm{E}}\right) $. To obtain
corresponding geometrical object $\mathcal{O}$ and corresponding relation $%
\mathcal{R}$ between the geometrical objects in other physical geometry $%
\mathcal{G}$, it is sufficient to replace the Euclidean world function $%
\sigma _{\mathrm{E}}$ by the world function $\sigma $ of the physical
geometry $\mathcal{G}$ in description of $\mathcal{O}_{\mathrm{E}}\left(
\sigma _{\mathrm{E}}\right) $ and $\mathcal{R}_{\mathrm{E}}\left( \sigma _{%
\mathrm{E}}\right) $. 
\[
\mathcal{O}_{\mathrm{E}}\left( \sigma _{\mathrm{E}}\right) \rightarrow 
\mathcal{O}_{\mathrm{E}}\left( \sigma \right) ,\qquad \mathcal{R}_{\mathrm{E}%
}\left( \sigma _{\mathrm{E}}\right) \rightarrow \mathcal{R}_{\mathrm{E}%
}\left( \sigma \right) 
\]
Index 'E' shows that the geometric object is constructed on the basis of the
Euclidean axiomatics. Thus, one can obtain another physical geometry $%
\mathcal{G}$ from the Euclidean geometry $\mathcal{G}_{\mathrm{E}}$ by a
simple replacement of $\sigma _{\mathrm{E}}$ by $\sigma $. For such a
construction one needs no axiomatics and no reasonings. One needs no means
of descriptions (topological structures, continuity, coordinate system,
manifold, dimension, etc.). In fact, one uses implicitly the axiomatics of
the Euclidean geometry, which is deformed by the replacement $\sigma _{%
\mathrm{E}}\rightarrow \sigma $. This replacement may be interpreted as a
deformation of the Euclidean space. Absence of a reference to the means of
description is an advantage of the considered method of the geometry
construction. Besides, there is no necessity to construct the whole geometry 
$\mathcal{G}$. We can construct and investigate only that part of the
geometry $\mathcal{G}$ which we are interested in. Any physical geometry may
be constructed as a result of a deformation of the Euclidean geometry.

The geometric object $\mathcal{O}$ is described by means of the
skeleton-envelope method \cite{R01}. It means that any geometric object $%
\mathcal{O}$ is considered to be a set of intersections and joins of
elementary geometric objects (EGO).

The finite set $\mathcal{P}^{n}\equiv \left\{ P_{0},P_{1},...,P_{n}\right\}
\subset \Omega $ of parameters of the envelope function $f_{\mathcal{P}^{n}}$
is the skeleton of elementary geometric object (EGO) $\mathcal{E}\subset
\Omega $. The set $\mathcal{E}\subset \Omega $ of points forming EGO is
called the envelope of its skeleton $\mathcal{P}^{n}$. For continuous
physical geometry the envelope $\mathcal{E}$ is usually a continual set of
points. The envelope function $f_{\mathcal{P}^{n}}$, determining EGO is a
function of the running point $R\in \Omega $ and of parameters $\mathcal{P}%
^{n}\subset \Omega $. The envelope function $f_{\mathcal{P}^{n}}$ is
supposed to be an algebraic function of $s$ arguments $w=\left\{
w_{1},w_{2},...w_{s}\right\} $, $s=(n+2)(n+1)/2$. Each of arguments $%
w_{k}=\sigma \left( Q_{k},L_{k}\right) $ is a $\sigma $-function of two
arguments $Q_{k},L_{k}\in \left\{ R,\mathcal{P}^{n}\right\} $, either
belonging to skeleton $\mathcal{P}^{n}$, or coinciding with the running
point $R$. Thus, any elementary geometric object $\mathcal{E}$ is determined
by its skeleton and its envelope function.

For instance, the sphere $\mathcal{S}(P_{0},P_{1})$ with the center at the
point $P_{0}$ is determined by the relation 
\begin{equation}
\mathcal{S}(P_{0},P_{1})=\left\{ R|f_{P_{0}P_{1}}\left( R\right) =0\right\}
,\qquad f_{P_{0}P_{1}}\left( R\right) =\sqrt{2\sigma \left(
P_{0},P_{1}\right) }-\sqrt{2\sigma \left( P_{0},R\right) }  \label{b1.6}
\end{equation}
where $P_{1}$ is a point belonging to the sphere. The elementary object $%
\mathcal{E}$ is determined in all physical geometries at once. In
particular, it is determined in the proper Euclidean geometry, where we can
obtain its meaning. We interpret the elementary geometrical object $\mathcal{%
E}$, using our knowledge of the proper Euclidean geometry. Thus, the proper
Euclidean geometry is used as a sample geometry for interpretation of any
physical geometry.

We do not try to repeat subscriptions of Euclid at construction of the
geometry. We take the geometrical objects and relations between them,
prepared in the framework of the Euclidean geometry and describe them in
terms of the world function. Thereafter we deform them, replacing the
Euclidean world function $\sigma _{\mathrm{E}}$ by the world function $%
\sigma $ of the geometry in question. In practice the construction of the
elementary geometry object is reduced to the representation of the
corresponding Euclidean geometrical object in the $\sigma $-immanent form,
i.e. in terms of the Euclidean world function. The last problem is the
problem of the proper Euclidean geometry. The problem of representation of
the geometrical object (or relation between objects) in the $\sigma $%
-immanent form is a real problem of the physical geometry construction.

It is very important, that such a construction does not use coordinates and
other methods of description, because the application of the means of
description imposes constraints on the constructed geometry. Any means of
description is a structure $St$ given on the basic Euclidean geometry with
the world function $\sigma _{\mathrm{E}}$. Replacement $\sigma _{\mathrm{E}%
}\rightarrow \sigma $ is sufficient for construction of unique physical
geometry $\mathcal{G}_{\sigma }$. If we use an additional structure $St$ for
construction of physical geometry, we obtain, in general, other geometry $%
\mathcal{G}_{St}$, which coincide with $\mathcal{G}_{\sigma }$ not for all $%
\sigma $, but only for some of world functions $\sigma $. Thus, a use of
additional means of description restricts the list of possible physical
geometries. For instance, if we use the coordinate description at
construction of the physical geometry, the obtained geometry appears to be
continuous, because description by means of the coordinates is effective
only for continuous geometries, where the number of coordinates coincides
with the geometry dimension.

Constructing geometry $\mathcal{G}$ by means of a deformation we use
essentially the fact that the proper Euclidean geometry $\mathcal{G}_{%
\mathrm{E}}$ is a mathematical geometry, which has been constructed on the
basis of Euclidean axiomatics and logical reasonings.

We shall refer to the described method of the physical geometry construction
as the deformation principle and interpret the deformation in the broad
sense of the word. In particular, a deformation of the Euclidean space may
transform an Euclidean surface into a point, and an Euclidean point into a
surface. Such a deformation may remove some points of the Euclidean space,
violating its continuity, or decreasing its dimension. Such a deformation
may add supplemental points to the Euclidean space, increasing its
dimension. In other words, the deformation principle is a very general
method of the physical geometry construction.

The deformation principle as a method of the physical geometry construction
contains two essential stages:

(i) Representation of geometrical objects $\mathcal{O}$ and relations $%
\mathcal{R}$ of the Euclidean geometry in the $\sigma $-immanent form, i.e.
in terms and only in terms of the world function $\sigma _{\mathrm{E}}$.

(ii) Replacement of the Euclidean world function $\sigma _{\mathrm{E}}$ by
the world function $\sigma $ of the geometry in question.

A physical geometry, constructed by means of the only deformation principle
(i.e. without a use of other methods of the geometry construction) is called
T-geometry (tubular geometry) \cite{R90,R01,R002}. The T-geometry is the
most general kind of the physical geometry.

Application of the deformation principle is restricted by two constraints.

1. Describing Euclidean geometric objects $\mathcal{O}\left( \sigma _{%
\mathrm{E}}\right) $ and Euclidean relation $\mathcal{R}\left( \sigma _{%
\mathrm{E}}\right) $ in terms of $\sigma _{\mathrm{E}}$, we are not to use
special properties of Euclidean world function $\sigma _{\mathrm{E}}$. In
particular, definitions of $\mathcal{O}\left( \sigma _{\mathrm{E}}\right) $
and $\mathcal{R}\left( \sigma _{\mathrm{E}}\right) $ are to have similar
form in Euclidean geometries of different dimensions. They must not depend
on the dimension of the Euclidean space.

2. The deformation principle is to be applied separately from other methods
of the geometry construction. In particular, one may not use topological
structures in construction of a physical geometry, because for effective
application of the deformation principle the obtained physical geometry must
be determined only by the world function (metric).

\section{Description of the proper Euclidean space in terms of the world
function}

The crucial point of the T-geometry construction is the description of the
proper Euclidean geometry in terms of the Euclidean world function $\sigma _{%
\mathrm{E}}$. We shall refer to this method of description as the $\sigma $%
-immanent description. Unfortunately, it was unknown for many years,
although all physicists knew that the infinitesimal interval $dS=\sqrt{%
g_{ik}dx^{i}dx^{k}}$ is the unique essential characteristic of the
space-time geometry, and changing this expression, we change the space-time
geometry. From physical viewpoint the $\sigma $-immanent description is very
reasonable, because it does not contain any extrinsic information. The $%
\sigma $-immanent description does not refer to the means of description
(dimension, manifold, coordinate system). Absence of references to means of
description is important in the relation, that there is no necessity to
separate the information on the geometry in itself from the information on
the means of description. The $\sigma $-immanent description contains only
essential characteristic of geometry: its world function. At first the $%
\sigma $-immanent description was obtained in 1990 \cite{R90}.

The first question concerning the $\sigma $-immanent description is as
follows. Does the world function contain sufficient information for
description of a physical geometry? The answer is affirmative, at least, in
the case of the proper Euclidean geometry, and this answer is given by the
prove of the following theorem.

Let $\sigma $-space $V=\left\{ \sigma ,\Omega \right\} $ be a set $\Omega $
of points $P$ with the given world function $\sigma $%
\begin{equation}
\sigma :\qquad \Omega \times \Omega \rightarrow \mathbf{R},\qquad \sigma
\left( P,P\right) =0,\qquad \forall P\in \Omega  \label{a1.3}
\end{equation}
where $\mathbf{R}$ denotes the set of all real numbers. Let the vector $%
\mathbf{P}_{0}\mathbf{P}_{1}\mathbf{=}\left\{ P_{0},P_{1}\right\} $ be the
ordered set of two points $P_{0}$, $P_{1}$, and its length $\left| \mathbf{P}%
_{0}\mathbf{P}_{1}\right| $ is defined by the relation $\left| \mathbf{P}_{0}%
\mathbf{P}_{1}\right| ^{2}=2\sigma \left( P_{0},P_{1}\right) $.

\textit{Theorem}

The $\sigma $-space $V=\left\{ \sigma ,\Omega \right\} $ is the $n$%
-dimensional proper Euclidean space, if and only if the world function $%
\sigma $ satisfies the following conditions, written in terms of the world
function $\sigma $.

I. Condition of symmetry: 
\begin{equation}
\sigma \left( P,Q\right) =\sigma \left( Q,P\right) ,\qquad \forall P,Q\in
\Omega  \label{a1.4}
\end{equation}

II. Definition of the dimension: 
\begin{equation}
\exists \mathcal{P}^{n}\equiv \left\{ P_{0},P_{1},...P_{n}\right\} ,\qquad
F_{n}\left( \mathcal{P}^{n}\right) \neq 0,\qquad F_{k}\left( {\Omega }%
^{k+1}\right) =0,\qquad k>n  \label{b10}
\end{equation}
where $F_{n}\left( \mathcal{P}^{n}\right) $ is the Gram's determinant 
\begin{equation}
F_{n}\left( \mathcal{P}^{n}\right) =\det \left| \left| \left( \mathbf{P}_{0}%
\mathbf{P}_{i}.\mathbf{P}_{0}\mathbf{P}_{k}\right) \right| \right| =\det
\left| \left| g_{ik}\left( \mathcal{P}^{n}\right) \right| \right| ,\qquad
i,k=1,2,...n  \label{b11}
\end{equation}
The scalar product $\left( \mathbf{P}_{0}\mathbf{P}_{1}.\mathbf{Q}_{0}%
\mathbf{Q}_{1}\right) $ of two vectors $\mathbf{P}_{0}\mathbf{P}_{1}$ and $%
\mathbf{Q}_{0}\mathbf{Q}_{1}$ is defined by the relation 
\begin{equation}
\left( \mathbf{P}_{0}\mathbf{P}_{1}.\mathbf{Q}_{0}\mathbf{Q}_{1}\right)
=\sigma \left( P_{0},Q_{1}\right) +\sigma \left( P_{1},Q_{0}\right) -\sigma
\left( P_{0},Q_{0}\right) -\sigma \left( P_{1},Q_{1}\right)  \label{b11a}
\end{equation}
Vectors $\mathbf{P}_{0}\mathbf{P}_{i}$, $\;i=1,2,...n$ are basic vectors of
the rectilinear coordinate system $K_{n}$ with the origin at the point $%
P_{0} $, and the metric tensors $g_{ik}\left( \mathcal{P}^{n}\right) $, $%
g^{ik}\left( \mathcal{P}^{n}\right) $, \ $i,k=1,2,...n$ in $K_{n}$ are
defined by the relations 
\begin{equation}
\sum\limits_{k=1}^{k=n}g^{ik}\left( \mathcal{P}^{n}\right) g_{lk}\left( 
\mathcal{P}^{n}\right) =\delta _{l}^{i},\qquad g_{il}\left( \mathcal{P}%
^{n}\right) =\left( \mathbf{P}_{0}\mathbf{P}_{i}.\mathbf{P}_{0}\mathbf{P}%
_{l}\right) ,\qquad i,l=1,2,...n  \label{a15b}
\end{equation}

III. Linear structure of the Euclidean space: 
\begin{equation}
\sigma \left( P,Q\right) =\frac{1}{2}\sum\limits_{i,k=1}^{i,k=n}g^{ik}\left( 
\mathcal{P}^{n}\right) \left( x_{i}\left( P\right) -x_{i}\left( Q\right)
\right) \left( x_{k}\left( P\right) -x_{k}\left( Q\right) \right) ,\qquad
\forall P,Q\in \Omega  \label{a15a}
\end{equation}
where coordinates $x_{i}\left( P\right) ,$ $i=1,2,...n$ of the point $P$ are
covariant coordinates of the vector $\mathbf{P}_{0}\mathbf{P}$, defined by
the relation 
\begin{equation}
x_{i}\left( P\right) =\left( \mathbf{P}_{0}\mathbf{P}_{i}.\mathbf{P}_{0}%
\mathbf{P}\right) ,\qquad i=1,2,...n  \label{b12}
\end{equation}

IV: The metric tensor matrix $g_{lk}\left( \mathcal{P}^{n}\right) $ has only
positive eigenvalues 
\begin{equation}
g_{k}>0,\qquad k=1,2,...,n  \label{a15c}
\end{equation}

V. The continuity condition: the system of equations 
\begin{equation}
\left( \mathbf{P}_{0}\mathbf{P}_{i}.\mathbf{P}_{0}\mathbf{P}\right)
=y_{i}\in \mathbf{R},\qquad i=1,2,...n  \label{b14}
\end{equation}
considered to be equations for determination of the point $P$ as a function
of coordinates $y=\left\{ y_{i}\right\} $,\ \ $i=1,2,...n$ has always one
and only one solution. Conditions II -- V contain a reference to the
dimension $n$ of the Euclidean space.

As far as the $\sigma $-immanent description of the proper Euclidean
geometry is possible, it is possible for any T-geometry, because any
geometrical object $\mathcal{O}$ and any relation $\mathcal{R}$ in the
physical geometry $\mathcal{G}$ is obtained from the corresponding
geometrical object $\mathcal{O}_{\mathrm{E}}$ and from the corresponding
relation $\mathcal{R}_{\mathrm{E}}$ in the proper Euclidean geometry $%
\mathcal{G}_{\mathrm{E}}$ by means of the replacement $\sigma _{\mathrm{E}%
}\rightarrow \sigma $ in description of $\mathcal{O}_{\mathrm{E}}$ an $%
\mathcal{R}_{\mathrm{E}}$. For such a replacement be possible, the
description of $\mathcal{O}_{\mathrm{E}}$ and $\mathcal{R}_{\mathrm{E}}$ is
not to refer to special properties of $\sigma _{\mathrm{E}}$, described by
conditions II -- V. A formal indicator of the conditions II -- V application
is a reference to the dimension $n$, because any of conditions II -- V
contains a reference to the dimension $n$ of the proper Euclidean space.

If nevertheless we use one of special properties II -- V of the Euclidean
space in the $\sigma $-immanent description of a geometrical object $%
\mathcal{O}$, or relation $\mathcal{R}$ , we refer to the dimension $n$ and,
ultimately, to the coordinate system, which is only a means of description.

Let us show this in the example of the determination of the straight in the $%
n$-dimensional Euclidean space. The straight $\mathcal{T}_{P_{0}Q}$ in the
proper Euclidean space is defined by two its points $P_{0}$ and $Q$ $%
\;\left( P_{0}\neq Q\right) $ as the set of points $R$ 
\begin{equation}
\mathcal{T}_{P_{0}Q}=\left\{ R\;|\;\mathbf{P}_{0}\mathbf{Q}||\mathbf{P}_{0}%
\mathbf{R}\right\}  \label{b15}
\end{equation}
where condition $\mathbf{P}_{0}\mathbf{Q}||\mathbf{P}_{0}\mathbf{R}$ means
that vectors $\mathbf{P}_{0}\mathbf{Q}$ and $\mathbf{P}_{0}\mathbf{R}$ are
collinear, i.e. the scalar product $\left( \mathbf{P}_{0}\mathbf{Q}.\mathbf{P%
}_{0}\mathbf{R}\right) $ of these two vectors satisfies the relation 
\begin{equation}
\mathbf{P}_{0}\mathbf{Q}||\mathbf{P}_{0}\mathbf{R:\qquad }\left( \mathbf{P}%
_{0}\mathbf{Q}.\mathbf{P}_{0}\mathbf{R}\right) ^{2}=\left( \mathbf{P}_{0}%
\mathbf{Q}.\mathbf{P}_{0}\mathbf{Q}\right) \left( \mathbf{P}_{0}\mathbf{R}.%
\mathbf{P}_{0}\mathbf{R}\right)  \label{b16}
\end{equation}
where the scalar product is defined by the relation (\ref{b11a}). Thus, the
straight line $\mathcal{T}_{P_{0}Q}$ is defined $\sigma $-immanently, i.e.
in terms of the world function $\sigma $. We shall use two different names
(straight and tube) for the geometric object $\mathcal{T}_{P_{0}Q}$. We
shall use the term ''straight'', when we want to stress that $\mathcal{T}%
_{P_{0}Q}$ is a result of deformation of the Euclidean straight. We shall
use the term ''tube'', when we want to stress that $\mathcal{T}_{P_{0}Q}$
may be a many-dimensional surface.

In the Euclidean geometry one can use another definition of collinearity.
Vectors $\mathbf{P}_{0}\mathbf{Q}$ and $\mathbf{P}_{0}\mathbf{R}$ are
collinear, if components of vectors $\mathbf{P}_{0}\mathbf{Q}$ and $\mathbf{P%
}_{0}\mathbf{R}$ in some coordinate system are proportional. For instance,
in the $n$-dimensional Euclidean space one can introduce rectangular
coordinate system, choosing $n+1$ points $\mathcal{P}^{n}=\left\{
P_{0},P_{1},...P_{n}\right\} $ and forming $n$ basic vectors $\mathbf{P}_{0}%
\mathbf{P}_{i}$, $i=1,2,...n$. Then the collinearity condition can be
written in the form of $n$ equations 
\begin{equation}
\mathbf{P}_{0}\mathbf{Q}||\mathbf{P}_{0}\mathbf{R:\qquad }\left( \mathbf{P}%
_{0}\mathbf{P}_{i}.\mathbf{P}_{0}\mathbf{Q}\right) =a\left( \mathbf{P}_{0}%
\mathbf{P}_{i}.\mathbf{P}_{0}\mathbf{R}\right) ,\qquad i=1,2,...n,
\label{b17}
\end{equation}
where $a$ is some real constant. Relations (\ref{b17}) are relations for
covariant components of vectors $\mathbf{P}_{0}\mathbf{Q}$ and $\mathbf{P}%
_{0}\mathbf{R}$ in the considered coordinate system with basic vectors $%
\mathbf{P}_{0}\mathbf{P}_{i}$, $i=1,2,...n$. Let points $\mathcal{P}^{n}$ be
chosen in such a way, that $\left( \mathbf{P}_{0}\mathbf{P}_{1}.\mathbf{P}%
_{0}\mathbf{Q}\right) \neq 0$. Then eliminating the parameter $a$ from
relations (\ref{b17}), we obtain $n-1$ independent relations, and the
geometrical object 
\begin{eqnarray}
\mathcal{T}_{Q\mathcal{P}^{n}} &=&\left\{ R\;|\;\mathbf{P}_{0}\mathbf{Q}||%
\mathbf{P}_{0}\mathbf{R}\right\} =\bigcap\limits_{i=2}^{i=n}\mathcal{S}_{i},
\label{c2.1} \\
\mathcal{S}_{i} &=&\left\{ R\left| \frac{\left( \mathbf{P}_{0}\mathbf{P}_{i}.%
\mathbf{P}_{0}\mathbf{Q}\right) }{\left( \mathbf{P}_{0}\mathbf{P}_{1}.%
\mathbf{P}_{0}\mathbf{Q}\right) }=\frac{\left( \mathbf{P}_{0}\mathbf{P}_{i}.%
\mathbf{P}_{0}\mathbf{R}\right) }{\left( \mathbf{P}_{0}\mathbf{P}_{1}.%
\mathbf{P}_{0}\mathbf{R}\right) }\right. \right\} ,\qquad i=2,3,...n
\label{c2.2}
\end{eqnarray}
defined according to (\ref{b17}), depends on $n+2$ points $Q,\mathcal{P}^{n}$%
. This geometrical object $\mathcal{T}_{Q\mathcal{P}^{n}}$ is defined $%
\sigma $-immanently. It is a complex, consisting of the straight line and
the coordinate system, represented by $n+1$ points $\mathcal{P}^{n}=\left\{
P_{0},P_{1},...P_{n}\right\} $. In the Euclidean space the dependence on the
choice of the coordinate system and on $n+1$ points $\mathcal{P}^{n}$
determining this system, is fictitious. The geometrical object $\mathcal{T}%
_{Q\mathcal{P}^{n}}$ depends only on two points $P_{0},Q$ and coincides with
the straight line $\mathcal{T}_{P_{0}Q}$. But at deformations of the
Euclidean space the geometrical objects $\mathcal{T}_{Q\mathcal{P}^{n}}$ and 
$\mathcal{T}_{P_{0}Q}$ are deformed differently. The points $%
P_{1},P_{2},...P_{n}$ cease to be fictitious in definition of $\mathcal{T}_{Q%
\mathcal{P}^{n}}$, and geometrical objects $\mathcal{T}_{Q\mathcal{P}^{n}}$
and $\mathcal{T}_{P_{0}Q}$ become to be different geometric objects, in
general. But being different, in general, they may coincide in some special
cases.

What of the two geometrical objects in the deformed geometry should be
interpreted as the straight line, passing through the points $P_{0}$ and $Q$
in the geometry $\mathcal{G}$? Of course, it is $\mathcal{T}_{P_{0}Q}$,
because its definition does not contain a reference to a coordinate system,
whereas definition of $\mathcal{T}_{Q\mathcal{P}^{n}}$ depends on the choice
of the coordinate system, represented by points $\mathcal{P}^{n}$. In
general, definitions of geometric objects and relations between them are not
to refer to the means of description.

But in the given case the geometrical object $\mathcal{T}_{P_{0}Q}$ is, in
general, $(n-1)$-dimensional surface, whereas $\mathcal{T}_{Q\mathcal{P}%
^{n}} $ is an intersection of $(n-1)\;\;$ $(n-1)$-dimensional surfaces, i.e. 
$\mathcal{T}_{Q\mathcal{P}^{n}}$ is, in general, a one-dimensional curve.
The one-dimensional curve $\mathcal{T}_{Q\mathcal{P}^{n}}$ corresponds
better to our ideas on the straight line, than the $(n-1)$-dimensional
surface $\mathcal{T}_{P_{0}Q}$. Nevertheless, in physical geometry $\mathcal{%
G}$ it is $\mathcal{T}_{P_{0}Q}$, that is an analog of the Euclidean
straight line.

It is very difficult to overcome our conventional idea that the Euclidean
straight line cannot be deformed into many-dimensional surface, and \textit{%
this idea has been prevent for years from construction of T-geometries}.
Practically one uses such physical geometries, where deformation of the
Euclidean space transforms the Euclidean straight lines into one-dimensional
lines. It means that one chooses such geometries, where geometrical objects $%
\mathcal{T}_{P_{0}Q}$ and $\mathcal{T}_{Q\mathcal{P}^{n}}$ coincide. 
\begin{equation}
\mathcal{T}_{P_{0}Q}=\mathcal{T}_{Q\mathcal{P}^{n}}  \label{b19}
\end{equation}
Condition (\ref{b19}) of coincidence of the objects $\mathcal{T}_{P_{0}Q}$
and $\mathcal{T}_{Q\mathcal{P}^{n}}$, imposed on the T-geometry, restricts
list of possible T-geometries.

Let us consider the metric geometry, given on the set $\Omega $ of points.
The metric space $M=\left\{ \rho ,\Omega \right\} $ is given by the metric
(distance) $\rho $. 
\begin{eqnarray}
\rho &:&\quad \Omega \times \Omega \rightarrow \lbrack 0,\infty )\subset 
\mathbf{R}  \label{c2.3} \\
\rho (P,P) &=&0,\qquad \rho (P,Q)=\rho (Q,P),\qquad \forall P,Q\in \Omega
\label{c2.4} \\
\rho (P,Q) &\geq &0,\qquad \rho (P,Q)=0,\quad \text{iff }P=Q,\qquad \forall
P,Q\in \Omega  \label{c2.5} \\
0 &\leq &\rho (P,R)+\rho (R,Q)-\rho (P,Q),\qquad \forall P,Q,R\in \Omega
\label{c2.6}
\end{eqnarray}
where $\mathbf{R}$ denotes the set of all real numbers. At first sight the
metric space is a special case of the $\sigma $-space (\ref{a1.3}), and the
metric geometry is a special case of the T-geometry with additional
constraints (\ref{c2.5}), (\ref{c2.6}) imposed on the world function $\sigma
=\frac{1}{2}\rho ^{2}$. However it is not so, because the metric geometry
does not use the deformation principle. The fact, that the Euclidean
geometry can be described $\sigma $-immanently, as well as the conditions (%
\ref{b10}) - (\ref{b14}), were not known until 1990. Additional (with
respect to the $\sigma $-space) constraints (\ref{c2.5}), (\ref{c2.6}) are
imposed to eliminate the situation, when the straight line is not a
one-dimensional line. The fact is that, in the metric geometry the shortest
(straight) line can be constructed only in the case, when it is
one-dimensional.

Let us consider the set $\mathcal{EL}\left( P,Q,a\right) $ of points $R$ 
\begin{equation}
\mathcal{EL}\left( P,Q,a\right) =\left\{ R|f_{P,Q,a}\left( R\right)
=0\right\} ,\qquad f_{P,Q,a}\left( R\right) =\rho (P,R)+\rho (R,Q)-2a
\label{c2.8}
\end{equation}
If the metric space coincides with the proper Euclidean space, this set of
points is an ellipsoid with focuses at the points $P,Q$ and the large
semiaxis $a$. The relations $f_{P,Q,a}\left( R\right) >0$, $f_{P,Q,a}\left(
R\right) =0$, $f_{P,Q,a}\left( R\right) <0$ determine respectively external
points, boundary points and internal points of the ellipsoid. If $\rho
\left( P,Q\right) =2a$, we obtain the degenerate ellipsoid, which coincides
with the segment $\mathcal{T}_{\left[ PQ\right] }$ of the straight line,
passing through the points $P$, $Q$.\ In the proper Euclidean geometry, the
degenerate ellipsoid is one-dimensional segment of the straight line, but it
is not evident that it is one-dimensional in the case of arbitrary metric
geometry. For such a degenerate ellipsoid be one-dimensional in the
arbitrary metric space, it is necessary that any degenerate ellipsoid $%
\mathcal{EL}\left( P,Q,\rho \left( P,Q\right) /2\right) $ have no internal
points. This constraint is written in the form 
\begin{equation}
f_{P,Q,\rho \left( P,Q\right) /2}\left( R\right) =\rho (P,R)+\rho (R,Q)-\rho
(P,Q)\geq 0  \label{c2.9}
\end{equation}

Comparing relation (\ref{c2.9}) with (\ref{c2.6}), we see that the
constraint (\ref{c2.6}) is introduced to make the straight (shortest) line
to be one-dimensional (absence of internal points in the geometrical object
determined by two points).

As far as the metric geometry does not use the deformation principle, it is
a poor geometry, because in the framework of this geometry one cannot
construct the scalar product of two vectors, define linear independence of
vectors and construct such geometrical objects as planes. All these objects
as well as other are constructed on the basis of the deformation of the
proper Euclidean geometry.

Generalizing the metric geometry, Menger \cite{M28} and Blumenthal \cite{B53}
removed the triangle axiom (\ref{c2.6}). They tried to construct the
distance geometry, which would be a more general geometry, than the metric
one. As far as they did not use the deformation principle, they could not
determine the shortest (straight) line without a reference to the
topological concept of the curve $\mathcal{L}$, defined as a continuous
mapping 
\begin{equation}
\mathcal{L}:\qquad \left[ 0,1\right] \rightarrow \Omega  \label{a1.1}
\end{equation}
which cannot be expressed only via the distance. As a result the distance
geometry appeared to be not a pure metric geometry, what the T-geometry is.

\section{Conditions of the deformation principle \newline
application}

Riemannian geometries satisfy the condition (\ref{b19}). The Riemannian
geometry is a kind of inhomogeneous physical geometry, and, hence, it uses
the deformation principle. Constructing the Riemannian geometry, the
infinitesimal Euclidean distance is deformed into the Riemannian distance.
The deformation is chosen in such a way that any Euclidean straight line $%
\mathcal{T}_{\mathrm{E}P_{0}Q}$, passing through the point $P_{0}$,
collinear to the vector $\mathbf{P}_{0}\mathbf{Q}$, transforms into the
geodesic $\mathcal{T}_{P_{0}Q}$, passing through the point $P_{0}$,
collinear to the vector $\mathbf{P}_{0}\mathbf{Q}$ in the Riemannian space.

Note that in T-geometries, satisfying the condition (\ref{b19}) for all
points $Q,\mathcal{P}^{n}$, the straight line 
\begin{equation}
\mathcal{T}_{Q_{0};P_{0}Q}=\left\{ R\;|\;\mathbf{P}_{0}\mathbf{Q}||\mathbf{Q}%
_{0}\mathbf{R}\right\}  \label{b3.0}
\end{equation}
passing through the point $Q_{0}$ collinear to the vector $\mathbf{P}_{0}%
\mathbf{Q}$, is not a one-dimensional line, in general. If the Riemannian
geometries be T-geometries, they would contain non-one-dimensional geodesics
(straight lines). But the Riemannian geometries are not T-geometries,
because at their construction one uses not only the deformation principle,
but some other methods, containing a reference to the means of description.
In particular, in the Riemannian geometries the absolute parallelism is
absent, and one cannot to define a straight line (\ref{b3.0}), because the
relation $\mathbf{P}_{0}\mathbf{Q}||\mathbf{Q}_{0}\mathbf{R}$ is not
defined, if points $P_{0}$ and $Q_{0}$ do not coincide. On one hand, a lack
of absolute parallelism allows one to go around the problem of
non-one-dimensional straight lines. On the other hand, it makes the
Riemannian geometries to be inconsistent, because they cease to be
T-geometries, which are consistent by the construction (see for details \cite
{R02}).

The fact is that the application of \textit{only deformation principle }is
sufficient for construction of a physical geometry. Besides, such a
construction is consistent, because the original Euclidean geometry is
consistent and, deforming it, we do not use any reasonings. If we introduce
additional structure (for instance, a topological structure) we obtain a
fortified physical geometry, i.e. a physical geometry with additional
structure on it. The physical geometry with additional structure on it is a
more pithy construction, than the physical geometry simply. But it is valid
only in the case, when we consider the additional structure as an addition
to the physical geometry. If we use an additional structure in construction
of the geometry, we identify the additional structure with one of structures
of the physical geometry. If we demand that the additional structure to be a
structure of physical geometry, we restrict an application of the
deformation principle and reduce the list of possible physical geometries,
because coincidence of the additional structure with some structure of a
physical geometry is possible not for all physical geometries, but only for
some of them.

Let, for instance, we use concept of a curve $\mathcal{L}$ (\ref{a1.1}) for
construction of a physical geometry. The concept of curve $\mathcal{L}$,
considered as a continuous mapping is a topological structure, which cannot
be expressed only via the distance or via the world function. A use of the
mapping (\ref{a1.1}) needs an introduction of topological space and, in
particular, the concept of continuity. If we identify the topological curve (%
\ref{a1.1}) with the ''metrical'' curve, defined as a broken line 
\begin{equation}
\mathcal{T}_{\mathrm{br}}=\bigcup\limits_{i}\mathcal{T}_{\left[ P_{i}P_{i+1}%
\right] },\qquad \mathcal{T}_{\left[ P_{i}P_{i+1}\right] }=\left\{ R|\sqrt{%
2\sigma \left( P_{i},P_{i+1}\right) }-\sqrt{2\sigma \left( P_{i},R\right) }-%
\sqrt{2\sigma \left( R,P_{i+1}\right) }\right\}  \label{a1.2}
\end{equation}
consisting of the straight line segments $\mathcal{T}_{\left[ P_{i}P_{i+1}%
\right] }$ between the points $P_{i}$, $P_{i+1}$, we truncate the list of
possible geometries, because such an identification is possible only in some
physical geometries. Identifying (\ref{a1.1}) and (\ref{a1.2}), we eliminate
all discrete physical geometries and those continuous physical geometries,
where the segment $\mathcal{T}_{\left[ P_{i}P_{i+1}\right] }$ of straight
line is a surface, but not a one-dimensional set of points. Thus, additional
structures may lead to (i) a fortified physical geometry, (ii) a restricted
physical geometry and (iii) a restricted fortified physical geometry. The
result depends on the method of the additional structure application.

Note that some constraints (continuity, convexity, lack of absolute
parallelism), imposed on physical geometries are a result of a disagreement
of the applied means of the geometry construction. In the T-geometry, which
uses only the deformation principle, there are no such restrictions.
Besides, the T-geometry accepts some new property of a physical geometry,
which is not accepted by conventional versions of physical geometry. This
property, called the geometry nondegeneracy, follows directly from the
application of arbitrary deformations to the proper Euclidean geometry.

The geometry is degenerate at the point $P_{0}$ in the direction of the
vector $\mathbf{Q}_{0}\mathbf{Q}$, $\left| \mathbf{Q}_{0}\mathbf{Q}\right|
\neq 0$, if the relations 
\begin{equation}
\mathbf{Q}_{0}\mathbf{Q}\uparrow \uparrow \mathbf{P}_{0}\mathbf{R:\qquad }%
\left( \mathbf{Q}_{0}\mathbf{Q}.\mathbf{P}_{0}\mathbf{R}\right) =\sqrt{%
\left| \mathbf{Q}_{0}\mathbf{Q}\right| \cdot \left| \mathbf{P}_{0}\mathbf{R}%
\right| },\qquad \left| \mathbf{P}_{0}\mathbf{R}\right| =a\neq 0
\label{b3.1}
\end{equation}
considered as equations for determination of the point $R$, have not more,
than one solution for any $a\neq 0$. Otherwise, the geometry is
nondegenerate at the point $P_{0}$ in the direction of the vector $\mathbf{Q}%
_{0}\mathbf{Q}$. Note that the first equation (\ref{b3.1}) is the condition
of the parallelism of vectors $\mathbf{Q}_{0}\mathbf{Q}$ and $\mathbf{P}_{0}%
\mathbf{R}$.

The proper Euclidean geometry is degenerate, i.e. it is degenerate at all
points in directions of all vectors. Considering the Minkowski geometry, one
should distinguish between the Minkowski T-geometry and Minkowski geometry.
The two geometries are described by the same world function and differ in
the definition of the parallelism. In the Minkowski T-geometry the
parallelism of two vectors $\mathbf{\mathbf{Q}_{0}\mathbf{Q}}$ and $\mathbf{%
\mathbf{P}_{0}\mathbf{R}}$ is defined by the first equation (\ref{b3.1}).
This definition is based on the deformation principle. In Minkowski geometry
the parallelism is defined by the relation of the type of (\ref{b17}) 
\begin{equation}
\mathbf{Q}_{0}\mathbf{Q}\uparrow \uparrow \mathbf{P}_{0}\mathbf{R:\qquad }%
\left( \mathbf{P}_{0}\mathbf{P}_{i}.\mathbf{Q}_{0}\mathbf{Q}\right) =a\left( 
\mathbf{P}_{0}\mathbf{P}_{i}.\mathbf{P}_{0}\mathbf{R}\right) ,\qquad
i=1,2,...n,\qquad a>0  \label{b3.1a}
\end{equation}
where points $\mathcal{P}^{n}=\left\{ P_{0},P_{1},...P_{n}\right\} $
determine a rectilinear coordinate system with basic vectors $\mathbf{P}_{0}%
\mathbf{P}_{i}$, $i=1,2,..n$ in the $n$-dimensional Minkowski geometry ($n$%
-dimensional pseudo-Euclidean geometry of index $1$). Dependence of the
definition (\ref{b3.1a}) on the points $\left( P_{1},P_{2},...P_{n}\right) $
is fictitious, but dependence on the number $n+1$ of points $\mathcal{P}^{n}$
is essential. Thus, definition (\ref{b3.1a}) depends on the method of the
geometry description.

The Minkowski T-geometry is degenerate at all points in direction of all
timelike vectors, and it is nondegenerate at all points in direction of all
spacelike vectors. The Minkowski geometry is degenerate at all points in
direction of all vectors. Conventionally one uses the Minkowski geometry,
ignoring the nondegeneracy in spacelike directions.

Considering the proper Riemannian geometry, one should distinguish between
the Riemannian T-geometry and the Riemannian geometry. The two geometries
are described by the same world function. They differ in the definition of
the parallelism. In the Riemannian T-geometry the parallelism of two vectors 
$\mathbf{\mathbf{Q}_{0}\mathbf{Q}}$ and $\mathbf{\mathbf{P}_{0}\mathbf{R}}$
is defined by the first equation (\ref{b3.1}). In the Riemannian geometry
the parallelism of two vectors $\mathbf{\mathbf{Q}_{0}\mathbf{Q}}$ and $%
\mathbf{\mathbf{P}_{0}\mathbf{R}}$ is defined only in the case, when the
points $P_{0}$ and $Q_{0}$ coincide. Parallelism of remote vectors $\mathbf{%
\mathbf{Q}_{0}\mathbf{Q}}$ and $\mathbf{\mathbf{P}_{0}\mathbf{R}}$ is not
defined, in general. This fact is known as absence of absolute parallelism.

The proper Riemannian T-geometry is locally degenerate, i.e. it is
degenerate at all points $P_{0}$ in direction of vectors $\mathbf{P}_{0}%
\mathbf{Q}$. In the general case, when $P_{0}\neq Q_{0}$, the proper
Riemannian T-geometry is nondegenerate, in general. The proper Riemannian
geometry is degenerate, because it is degenerate locally, whereas the
nonlocal degeneracy is not defined in the Riemannian geometry, because of
the lack of absolute parallelism. Conventionally one uses the Riemannian
geometry (not Riemannian T-geometry) and ignores the property of the
nondegeneracy completely.

From the viewpoint of the conventional approach to the physical geometry the
nondegeneracy is an undesirable property of a physical geometry, although
from the logical viewpoint and from viewpoint of the deformation principle
the nondegeneracy is an inherent property of a physical geometry. The
nonlocal nondegeneracy is ejected from the proper Riemannian geometry by
denial of existence of the remote vector parallelism. Nondegeneracy in the
spacelike directions is ejected from the Minkowski geometry by means of the
redefinition of the two vectors parallelism. To appreciate this, let us
consider an example.

\section{Simple example of nondegenerate space-time geometry}

The T-geometry \cite{R01} is defined on the $\sigma $-space $V=\left\{
\sigma ,\Omega \right\} $, where $\Omega $ is an arbitrary set of points and
the world function $\sigma $ is defined by the relations 
\begin{equation}
\sigma :\qquad \Omega \times \Omega \rightarrow \mathbf{R},\qquad \sigma
\left( P,Q\right) =\sigma \left( Q,P\right) ,\qquad \sigma \left( P,P\right)
=0,\qquad \forall P,Q\in \Omega  \label{b3.2}
\end{equation}
where $\mathbf{R}$ denotes the set of all real numbers. Geometrical objects
(vector $\mathbf{PQ}$, scalar product of vectors $\left( \mathbf{P}_{0}%
\mathbf{P}_{1}.\mathbf{Q}_{0}\mathbf{Q}_{1}\right) $, collinearity of
vectors $\mathbf{\ P}_{0}\mathbf{P}_{1}||\mathbf{Q}_{0}\mathbf{Q}_{1}$,
segment of straight line $\mathcal{T}_{\left[ P_{0}P_{1}\right] }$, etc.)
are defined on the $\sigma $-space in the same way, as they are defined $%
\sigma $-immanently in the proper Euclidean space. Practically one uses the
deformation principle, although it is not mentioned in all definitions.

Let us consider a simple example of the space-time geometry $\mathcal{G}_{%
\mathrm{d}}$, described by the T-geometry on 4-dimensional manifold $%
\mathcal{M}_{1+3}$. The world function $\sigma _{\mathrm{d}}$ is described
by the relation 
\begin{equation}
\sigma _{\mathrm{d}}=\sigma _{\mathrm{M}}+D\left( \sigma _{\mathrm{M}%
}\right) =\left\{ 
\begin{array}{ll}
\sigma _{\mathrm{M}}+d & \text{if\ }\sigma _{0}<\sigma _{\mathrm{M}} \\ 
\left( 1+\frac{d}{\sigma _{0}}\right) \sigma _{\mathrm{M}} & \text{if\ }%
0\leq \sigma _{\mathrm{M}}\leq \sigma _{0} \\ 
\sigma _{\mathrm{M}} & \text{if\ }\sigma _{\mathrm{M}}<0
\end{array}
\right.  \label{b3.3}
\end{equation}
where $d\geq 0$ and $\sigma _{0}>0$ are some constants. The quantity $\sigma
_{\mathrm{M}}$ is the world function in the Minkowski space-time geometry $%
\mathcal{G}_{\mathrm{M}}$. In the orthogonal rectilinear (inertial)
coordinate system $x=\left( t,\mathbf{x}\right) $ the world function $\sigma
_{\mathrm{M}}$ has the form 
\begin{equation}
\sigma _{\mathrm{M}}\left( x,x^{\prime }\right) =\frac{1}{2}\left(
c^{2}\left( t-t^{\prime }\right) ^{2}-\left( \mathbf{x}-\mathbf{x}^{\prime
}\right) ^{2}\right)  \label{b3.4}
\end{equation}
where $c$ is the speed of the light.

Let us compare the broken line (\ref{a1.2}) in Minkowski space-time geometry 
$\mathcal{G}_{\mathrm{M}}$ and in the distorted geometry $\mathcal{G}_{%
\mathrm{d}}$. We suppose that $\mathcal{T}_{\mathrm{br}}$ is timelike broken
line, and all links $\mathcal{T}_{\left[ P_{i}P_{i+1}\right] }$ of $\mathcal{%
T}_{\mathrm{br}}$ are timelike and have the same length 
\begin{equation}
\left| \mathbf{P}_{i}\mathbf{P}_{i+1}\right| _{\mathrm{d}}=\sqrt{2\sigma _{%
\mathrm{d}}\left( P_{i},P_{i+1}\right) }=\mu _{\mathrm{d}}>0,\qquad i=0,\pm
1,\pm 2,...  \label{b3.5}
\end{equation}
where indices ''d'' and ''M'' mean that the quantity is calculated by means
of $\sigma _{\mathrm{d}}$ and $\sigma _{\mathrm{M}}$ respectively. Vector $%
\mathbf{P}_{i}\mathbf{P}_{i+1}$ is regarded as the momentum of the particle
at the segment $\mathcal{T}_{\left[ P_{i}P_{i+1}\right] }$, devided by the
speed of the light $c$ (we take for simplicity that $c=1$). The quantity $%
\left| \mathbf{P}_{i}\mathbf{P}_{i+1}\right| =\mu $ is interpreted as its
(geometric) mass. It follows from definition (\ref{b11a}) and relation (\ref
{b3.3}), that for timelike vectors $\mathbf{P}_{i}\mathbf{P}_{i+1}$ with $%
\mu >\sqrt{2\sigma _{0}}$%
\begin{equation}
\left| \mathbf{P}_{i}\mathbf{P}_{i+1}\right| _{\mathrm{d}}^{2}=\mu _{\mathrm{%
d}}^{2}=\mu _{\mathrm{M}}^{2}+2d,\qquad \mu _{\mathrm{M}}^{2}>2\sigma _{0}
\label{b3.6}
\end{equation}
\begin{equation}
\left( \mathbf{P}_{i-1}\mathbf{P}_{i}.\mathbf{P}_{i}\mathbf{P}_{i+1}\right)
_{\mathrm{d}}=\left( \mathbf{P}_{i-1}\mathbf{P}_{i}.\mathbf{P}_{i}\mathbf{P}%
_{i+1}\right) _{\mathrm{M}}+d  \label{b3.7}
\end{equation}
Calculation of the shape of the segment $\mathcal{T}_{\left[ P_{0}P_{1}%
\right] }\left( \sigma _{\mathrm{d}}\right) $ in $\mathcal{G}_{\mathrm{d}}$
gives the relation 
\begin{equation}
r^{2}(\tau )=\left\{ 
\begin{array}{ll}
\tau ^{2}\mu _{\mathrm{d}}^{2}\frac{\left( 1-\frac{\tau d}{2\left( \sigma
_{0}+d\right) }\right) ^{2}}{\left( 1-\frac{2d}{\mu _{\mathrm{d}}^{2}}%
\right) }-\frac{\tau ^{2}\mu _{\mathrm{d}}^{2}\sigma _{0}}{\left( \sigma
_{0}+d\right) }, & 0<\tau <\frac{\sqrt{2(\sigma _{0}+d)}}{\mu _{\mathrm{d}}}
\\ 
\frac{3d}{2}+2d\left( \tau -1/2\right) ^{2}\left( 1-\frac{2d}{\mu _{\mathrm{d%
}}^{2}}\right) ^{-1}, & \frac{\sqrt{2(\sigma _{0}+d)}}{\mu _{\mathrm{d}}}%
<\tau <1-\frac{\sqrt{2(\sigma _{0}+d)}}{\mu _{\mathrm{d}}} \\ 
\left( 1-\tau \right) ^{2}\mu _{\mathrm{d}}^{2}\left[ \frac{\left( 1-\frac{%
\left( 1-\tau \right) d}{2\left( \sigma _{0}+d\right) }\right) ^{2}}{\left(
1-\frac{2d}{\mu _{\mathrm{d}}^{2}}\right) }-\frac{\sigma _{0}}{\left( \sigma
_{0}+d\right) }\right] , & 1-\frac{\sqrt{2(\sigma _{0}+d)}}{\mu _{\mathrm{d}}%
}<\tau <1
\end{array}
\right. ,  \label{b3.7a}
\end{equation}
where $r\left( \tau \right) $ is the spatial radius of the segment $\mathcal{%
T}_{\left[ P_{0}P_{1}\right] }\left( \sigma _{\mathrm{d}}\right) $ in the
coordinate system, where points $P_{0}$ and $P_{1}$ have coordinates $%
P_{0}=\left\{ 0,0,0,0\right\} $, $P_{1}=\left\{ \mu _{\mathrm{d}%
},0,0,0\right\} $ and $\tau $ is a parameter along the segment $\mathcal{T}_{%
\left[ P_{0}P_{1}\right] }\left( \sigma _{\mathrm{d}}\right) $ ($\tau \left(
P_{0}\right) =0$, $\tau \left( P_{1}\right) =1$). One can see from (\ref
{b3.7a}) that the characteristic value of the segment radius is $\sqrt{d}$.

Let the broken tube $\mathcal{T}_{\mathrm{br}}$ describe the ''world line''
of a free particle. It means by definition that any link $\mathbf{P}_{i-1}%
\mathbf{P}_{i}$ is parallel to the adjacent link $\mathbf{P}_{i}\mathbf{P}%
_{i+1}$%
\begin{equation}
\mathbf{P}_{i-1}\mathbf{P}_{i}\uparrow \uparrow \mathbf{P}_{i}\mathbf{P}%
_{i+1}:\qquad \left( \mathbf{P}_{i-1}\mathbf{P}_{i}.\mathbf{P}_{i}\mathbf{P}%
_{i+1}\right) -\left| \mathbf{P}_{i-1}\mathbf{P}_{i}\right| \cdot \left| 
\mathbf{P}_{i}\mathbf{P}_{i+1}\right| =0  \label{b3.8}
\end{equation}
Definition of parallelism is different in geometries $\mathcal{G}_{\mathrm{M}%
}$ and $\mathcal{G}_{\mathrm{d}}$. As a result links, which are parallel in
the geometry $\mathcal{G}_{\mathrm{M}}$, are not parallel in $\mathcal{G}_{%
\mathrm{d}}$ and vice versa.

Let $\mathcal{T}_{\mathrm{br}}\left( \sigma _{\mathrm{M}}\right) $ describe
the world line of a free particle in the geometry $\mathcal{G}_{\mathrm{M}}$%
. The angle $\vartheta _{\mathrm{M}}$ between the adjacent links in $%
\mathcal{G}_{\mathrm{M}}$ is defined by the relation 
\begin{equation}
\cosh \vartheta _{\mathrm{M}}=\frac{\left( \mathbf{P}_{-1}\mathbf{P}_{0}.%
\mathbf{P}_{0}\mathbf{P}_{1}\right) _{\mathrm{M}}}{\left| \mathbf{P}_{0}%
\mathbf{P}_{1}\right| _{\mathrm{M}}\cdot \left| \mathbf{P}_{-1}\mathbf{P}%
_{0}\right| _{\mathrm{M}}}=1  \label{b3.9}
\end{equation}
The angle $\vartheta _{\mathrm{M}}=0$, and the geometrical object $\mathcal{T%
}_{\mathrm{br}}\left( \sigma _{\mathrm{M}}\right) $ is a timelike straight
line on the manifold $\mathcal{M}_{1+3}$.

Let now $\mathcal{T}_{\mathrm{br}}\left( \sigma _{\mathrm{d}}\right) $
describe the world line of a free particle in the geometry $\mathcal{G}_{%
\mathrm{d}}$. The angle $\vartheta _{\mathrm{d}}$ between the adjacent links
in $\mathcal{G}_{\mathrm{d}}$ is defined by the relation 
\begin{equation}
\cosh \vartheta _{\mathrm{d}}=\frac{\left( \mathbf{P}_{i-1}\mathbf{P}_{i}.%
\mathbf{P}_{i}\mathbf{P}_{i+1}\right) _{\mathrm{d}}}{\left| \mathbf{P}_{i}%
\mathbf{P}_{i+1}\right| _{\mathrm{d}}\cdot \left| \mathbf{P}_{i-1}\mathbf{P}%
_{i}\right| _{\mathrm{d}}}=1  \label{b3.10}
\end{equation}
The angle $\vartheta _{\mathrm{d}}=0$ also. If we draw the broken tube $%
\mathcal{T}_{\mathrm{br}}\left( \sigma _{\mathrm{d}}\right) $ on the
manifold $\mathcal{M}_{1+3}$, using coordinates of basic points $P_{i}$ and
measure the angle $\vartheta _{\mathrm{dM}}$ between the adjacent links in
the Minkowski geometry $\mathcal{G}_{\mathrm{M}}$, we obtain for the angle $%
\vartheta _{\mathrm{dM}}$ the following relation 
\begin{equation}
\cosh \vartheta _{\mathrm{dM}}=\frac{\left( \mathbf{P}_{i-1}\mathbf{P}_{i}.%
\mathbf{P}_{i}\mathbf{P}_{i+1}\right) _{\mathrm{M}}}{\left| \mathbf{P}_{i}%
\mathbf{P}_{i+1}\right| _{\mathrm{M}}\cdot \left| \mathbf{P}_{i-1}\mathbf{P}%
_{i}\right| _{\mathrm{M}}}=\frac{\left( \mathbf{P}_{i-1}\mathbf{P}_{i}.%
\mathbf{P}_{i}\mathbf{P}_{i+1}\right) _{\mathrm{d}}-d}{\left| \mathbf{P}_{i}%
\mathbf{P}_{i+1}\right| _{\mathrm{d}}^{2}-2d}  \label{b3.11}
\end{equation}
Substituting the value of $\left( \mathbf{P}_{i-1}\mathbf{P}_{i}.\mathbf{P}%
_{i}\mathbf{P}_{i+1}\right) _{\mathrm{d}}$, taken from (\ref{b3.10}), we
obtain for the case, when $d\ll \mu _{\mathrm{d}}^{2}$ 
\begin{equation}
\cosh \vartheta _{\mathrm{dM}}=\frac{\mu _{\mathrm{d}}^{d}-d}{\mu _{\mathrm{d%
}}^{2}-2d}\approx 1+\frac{d}{\mu _{\mathrm{d}}^{2}},\qquad d\ll \mu _{%
\mathrm{d}}^{2}  \label{b3.12}
\end{equation}
Hence, $\vartheta _{\mathrm{dM}}\approx \sqrt{2d}/\mu _{\mathrm{d}}$. It
means, that the adjacent link is located on the cone of angle $\sqrt{2d}/\mu
_{\mathrm{d}}$, and the whole line $\mathcal{T}_{\mathrm{br}}\left( \sigma _{%
\mathrm{d}}\right) $ has a random shape, because any link wobbles with the
characteristic angle $\sqrt{2d}/\mu _{\mathrm{d}}$. The wobble angle depends
on the space-time distortion $d$ and on the particle mass $\mu _{\mathrm{d}}$%
. The wobble angle is small for the large mass of a particle. The random
displacement of the segment end is of the order $\mu _{\mathrm{d}}\vartheta
_{\mathrm{dM}}=\sqrt{2d}$, i.e. of the same order as the segment width. It
is reasonable, because these two phenomena have the common source: the
space-time distortion $D$.

One should note that the space-time geometry influences the stochasticity of
particle motion nonlocally in the sense, that the form of the world function
(\ref{b3.3}) for values of $\sigma _{\mathrm{M}}<\frac{1}{2}\mu _{\mathrm{d}%
}^{2}$ is unessential for the motion stochasticity of the particle of the
mass $\mu _{\mathrm{d}}$.

Such a situation, when the world line of a free particle is stochastic in
the deterministic geometry, and this stochasticity depends on the particle
mass, seems to be rather exotic and incredible. But experiments show that
the motion of real particles of small mass is stochastic indeed, and this
stochasticity increases, when the particle mass decreases. From physical
viewpoint a theoretical foundation of the stochasticity is desirable, and
some researchers invent stochastic geometries, noncommutative geometries and
other exotic geometrical constructions, to obtain the quantum stochasticity.
But in the Riemannian space-time geometry the particle motion does not
depend on the particle mass, and in the framework of the Riemannian
space-time geometry it is difficult to explain the quantum stochasticity by
the space-time geometry properties. Distorted geometry $\mathcal{G}_{\mathrm{%
d}}$ explains the stochasticity and its dependence on the particle mass
freely. Besides, at proper choice of the distortion $d$ the statistical
description of stochastic $\mathcal{T}_{\mathrm{br}}$ leads to the quantum
description (Schr\"{o}dinger equation) \cite{R91}. It is sufficient to set 
\begin{equation}
d=\frac{\hbar }{2bc},  \label{b3.13}
\end{equation}
where $\hbar $ is the quantum constant, $c$ is the speed of the light, and $%
b $ is some universal constant, connecting the geometrical mass $\mu $ with
the usual particle mass $m$ by means of the relation 
\begin{equation}
m=b\mu  \label{b3.14}
\end{equation}
In other words, the distorted space-time geometry (\ref{b3.3}) is closer to
the real space-time geometry, than the Minkowski geometry $\mathcal{G}_{%
\mathrm{M}}$.

\section{Statistical description of stochastic world tubes}

Statistical description of world lines cannot be a probabilistic statistical
description, because the number of world lines may be negative. Indeed, the
density of world lines in the vicinity of the space-time point $x$ is
defined by the relation 
\begin{equation}
dN=j^{k}dS_{k}  \label{b4.1}
\end{equation}
where $dN$ is the flux of world lines through the spacelike 3-area $dS_{k}$.
The 4-vector $j^{k}=j^{k}\left( x\right) $ describes the world-lines density
in the vicinity of the point $x$. The quantity $dN$ may be interpreted as
the number of world lines in the vicinity of the point $x$. This number may
be negative.

In the nonrelativistic case the relation (\ref{b4.1}) turns into the
relation 
\begin{equation}
dN=j^{0}dS_{0}=\rho dV  \label{b4.2}
\end{equation}
where the particle density $j^{0}=\rho \geq 0$, and $\rho $ may be a ground
for introduction of the probability density. In the relativistic case one
cannot introduce the probability density, because the world line density is
described by the 4-vector $j^{k}$.

For statistical description of stochastic world lines we use the dynamical
conception of statistical description (DCSD), which does not use the concept
of the probability \cite{R2002}.

Let $\mathcal{S}_{\mathrm{st}}$ be stochastic particle, whose state $X$ is
described by variables $\left\{ \mathbf{x},\frac{d\mathbf{x}}{dt}\right\} $,
where $\mathbf{x}$ is the particle position. Evolution of the particle state
is stochastic, and there exist no dynamic equations for $\mathcal{S}_{%
\mathrm{st}}$. Evolution of the state of $\mathcal{S}_{\mathrm{st}}$
contains both regular and stochastic components. To separate the regular
evolution components, we consider a set (statistical ensemble) $\mathcal{E}%
\left[ \mathcal{S}_{\mathrm{st}}\right] $ of many independent identical
stochastic particles $\mathcal{S}_{\mathrm{st}}$. All stochastic particles $%
\mathcal{S}_{\mathrm{st}}$ start from the same initial state. It means that
all $\mathcal{S}_{\mathrm{st}}$ are prepared in the same way. If the number $%
N$ of $\mathcal{S}_{\mathrm{st}}$ is very large, the stochastic elements of
evolution compensate each other, but regular ones are accumulated. In the
limit $N\rightarrow \infty $ the statistical ensemble $\mathcal{E}\left[ 
\mathcal{S}_{\mathrm{st}}\right] $ turns into a dynamic system, whose state
evolves according to some dynamic equations.

Let the statistical ensemble $\mathcal{E}_{\mathrm{d}}\left[ \mathcal{S}_{%
\mathrm{d}}\right] $ of deterministic classical particles $\mathcal{S}_{%
\mathrm{d}}$ be described by the action $\mathcal{A}_{\mathcal{E}_{\mathrm{d}%
}\left[ \mathcal{S}_{\mathrm{d}}\left( P\right) \right] }$, where $P$ are
parameters describing $\mathcal{S}_{\mathrm{d}}$ (for instance, mass,
charge). Let under influence of some stochastic agent the deterministic
particle $\mathcal{S}_{\mathrm{d}}$ turn into a stochastic particle $%
\mathcal{S}_{\mathrm{st}}$. The action $\mathcal{A}_{\mathcal{E}_{\mathrm{st}%
}\left[ \mathcal{S}_{\mathrm{st}}\right] }$ for the statistical ensemble $%
\mathcal{E}_{\mathrm{st}}\left[ \mathcal{S}_{\mathrm{st}}\right] $ is
reduced to the action $\mathcal{A}_{\mathcal{S}_{\mathrm{red}}\left[ 
\mathcal{S}_{\mathrm{d}}\right] }=\mathcal{A}_{\mathcal{E}_{\mathrm{st}}%
\left[ \mathcal{S}_{\mathrm{st}}\right] }$ for some set $\mathcal{S}_{%
\mathrm{red}}\left[ \mathcal{S}_{\mathrm{d}}\right] $ of identical
interacting deterministic particles $\mathcal{S}_{\mathrm{d}}$. The action $%
\mathcal{A}_{\mathcal{S}_{\mathrm{red}}\left[ \mathcal{S}_{\mathrm{d}}\right]
}$ as a functional of $\mathcal{S}_{\mathrm{d}}$ has the form $\mathcal{A}_{%
\mathcal{E}_{\mathrm{d}}\left[ \mathcal{S}_{\mathrm{d}}\left( P_{\mathrm{eff}%
}\right) \right] }$, where parameters $P_{\mathrm{eff}}$ are parameters $P$
of the deterministic particle $\mathcal{S}_{\mathrm{d}}$, averaged over the
statistical ensemble, and this averaging describes interaction of particles $%
\mathcal{S}_{\mathrm{d}}$ in the set $\mathcal{S}_{\mathrm{red}}\left[ 
\mathcal{S}_{\mathrm{d}}\right] $. It means that 
\begin{equation}
\mathcal{A}_{\mathcal{E}_{\mathrm{st}}\left[ \mathcal{S}_{\mathrm{st}}\right]
}=\mathcal{A}_{\mathcal{S}_{\mathrm{red}}\left[ \mathcal{S}_{\mathrm{d}%
}\left( P\right) \right] }=\mathcal{A}_{\mathcal{E}_{\mathrm{d}}\left[ 
\mathcal{S}_{\mathrm{d}}\left( P_{\mathrm{eff}}\right) \right] }
\label{a0.6a}
\end{equation}
In other words, stochasticity of particles $\mathcal{S}_{\mathrm{st}}$ in
the ensemble $\mathcal{E}_{\mathrm{st}}\left[ \mathcal{S}_{\mathrm{st}}%
\right] $ is replaced by interaction of $\mathcal{S}_{\mathrm{d}}$ in $%
\mathcal{S}_{\mathrm{red}}\left[ \mathcal{S}_{\mathrm{d}}\right] $, and this
interaction is described by a change 
\begin{equation}
P\rightarrow P_{\mathrm{eff}}  \label{a0.6b}
\end{equation}
in the action $\mathcal{A}_{\mathcal{E}_{\mathrm{d}}\left[ \mathcal{S}_{%
\mathrm{d}}\left( P\right) \right] }$.

The free particle has the unique parameter - its mass $m$, and the action $%
\mathcal{A}_{\mathcal{S}_{\mathrm{d}}}$ for the free deterministic particle
has the form 
\begin{equation}
\mathcal{S}_{\mathrm{d}}:\qquad \mathcal{A}_{\mathcal{S}_{\mathrm{d}}}\left[ 
\mathbf{x}\right] =\int \frac{m}{2}\left( \frac{d\mathbf{x}}{dt}\right)
^{2}dt  \label{c4.3}
\end{equation}
where $\mathbf{x}=\mathbf{x}\left( t\right) =\left\{ x^{1}\left( t\right)
,x^{2}\left( t\right) ,x^{3}\left( t\right) \right\} $, and the time $t$ is
the independent variable.

The action $\mathcal{A}_{\mathcal{E}_{\mathrm{d}}\left[ \mathcal{S}_{\mathrm{%
d}}\left( P\right) \right] }$ for the pure statistical ensemble $\mathcal{E}%
_{\mathrm{d}}\left[ \mathcal{S}_{\mathrm{d}}\right] $ of free deterministic
particles $\mathcal{S}_{\mathrm{d}}$ has the form 
\begin{equation}
\mathcal{E}_{\mathrm{d}}\left[ \mathcal{S}_{\mathrm{d}}\right] :\qquad 
\mathcal{A}_{\mathcal{E}_{\mathrm{d}}\left[ \mathcal{S}_{\mathrm{d}}\right] }%
\left[ \mathbf{x}\right] =\int \frac{m}{2}\left( \frac{d\mathbf{x}}{dt}%
\right) ^{2}dtd\mathbf{\xi }  \label{c4.4}
\end{equation}
where $\mathbf{x}=\mathbf{x}\left( t,\mathbf{\xi }\right) =\left\{
x^{1}\left( t,\mathbf{\xi }\right) ,x^{2}\left( t,\mathbf{\xi }\right)
,x^{3}\left( t,\mathbf{\xi }\right) \right\} $. Independent variables $%
\mathbf{\xi =}\left\{ \xi _{1},\xi _{2},\xi _{3}\right\} $ label elements $%
\mathcal{S}_{\mathrm{d}}$ of the statistical ensemble $\mathcal{E}_{\mathrm{d%
}}\left[ \mathcal{S}_{\mathrm{d}}\right] $. The variables $\mathbf{\xi }$
are known as Lagrangian coordinates. Statistical ensemble $\mathcal{E}_{%
\mathrm{d}}\left[ \mathcal{S}_{\mathrm{d}}\right] $ is a continuous dynamic
system, having infinite number of the freedom degrees, whereas the particle $%
\mathcal{S}_{\mathrm{d}}$ is the discrete dynamic system having six degrees
of freedom.

If the particles are stochastic, the action $\mathcal{A}_{\mathcal{E}_{%
\mathrm{st}}\left[ \mathcal{S}_{\mathrm{st}}\right] }$ for the pure
statistical ensemble $\mathcal{E}_{\mathrm{st}}\left[ \mathcal{S}_{\mathrm{st%
}}\right] $ of free quantum stochastic particles $\mathcal{S}_{\mathrm{st}}$
has the form

\begin{equation}
\mathcal{E}_{\mathrm{st}}\left[ \mathcal{S}_{\mathrm{st}}\right] :\qquad 
\mathcal{A}_{\mathcal{E}_{\mathrm{st}}\left[ \mathcal{S}_{\mathrm{st}}\right]
}\left[ \mathbf{x,u}\right] =\int \left\{ \frac{m}{2}\left( \frac{d\mathbf{x}%
}{dt}\right) ^{2}+\frac{m}{2}\mathbf{u}^{2}-\frac{\hbar }{2}\mathbf{\nabla u}%
\right\} dtd\mathbf{\xi }  \label{c4.5}
\end{equation}
where $\mathbf{u}=\mathbf{u}\left( t,\mathbf{x}\right) $ is a vector
function of arguments $t,\mathbf{x}$ (not of $t,\mathbf{\xi }$), and $%
\mathbf{x}=\mathbf{x}\left( t,\mathbf{\xi }\right) $ is a vector function of
independent variables $t,\mathbf{\xi }$. The 3-vector $\mathbf{u}$ describes
the mean value of the stochastic component of the particle motion, which is
a function of the variables $t,\mathbf{x}$. The first term $\frac{m}{2}%
\left( \frac{d\mathbf{x}}{dt}\right) ^{2}$ describes the energy of the
regular component of the stochastic particle motion. The second term $m%
\mathbf{u}^{2}/2$ describes the energy of the random component of velocity.
The components $\frac{d\mathbf{x}}{dt}$ and $\mathbf{u}$ of the total
velocity are connected with different degrees of freedom, and their energies
should be added in the expression for the Lagrange function density. The
last term $-\hbar \mathbf{\nabla u}/2$ describes interaction between the
regular component $\frac{d\mathbf{x}}{dt}$ and the random one $\mathbf{u}$.
Note that $m\mathbf{u}^{2}/2$ is a function of $t,\mathbf{x}$\textbf{. }It
influences on the regular component $\frac{d\mathbf{x}}{dt}$ as a potential
energy $U\left( t,\mathbf{x},\nabla \mathbf{x}\right) =-m\mathbf{u}^{2}/2$,
generated by the random component.

The dynamic system (\ref{c4.5}) is a statistical ensemble, because the
Lagrange function density of the action (\ref{c4.5}) does not depend on $%
\mathbf{\xi }$ explicitly, and we can represent the action for the single
system $\mathcal{S}_{\mathrm{st}}$%
\begin{equation}
\mathcal{S}_{\mathrm{st}}:\qquad \mathcal{A}_{\mathcal{S}_{\mathrm{st}}}%
\left[ \mathbf{x},\mathbf{u}\right] =\int \left\{ \frac{m}{2}\left( \frac{d%
\mathbf{x}}{dt}\right) ^{2}+\frac{m}{2}\mathbf{u}^{2}-\frac{\hbar }{2}%
\mathbf{\nabla u}\right\} dt  \label{b3.1c}
\end{equation}
Unfortunately, the expression for the action (\ref{b3.1c}) is only symbolic,
because the differential operator $\mathbf{\nabla }=\left\{ \partial
/\partial x^{\alpha }\right\} $, $\alpha =1,2,3$ is defined in the
continuous vicinity of the point $\mathbf{x}$, but not only for one point $%
\mathbf{x}$. The expression (\ref{b3.1c}) ceases to be symbolic, only if $%
\hbar =0$. In this case the last term, containing $\mathbf{\nabla }$
vanishes. Variation of (\ref{b3.1c}) with respect to $\mathbf{u}$ gives $%
\mathbf{u}=0$, and the action (\ref{b3.1c}) coincides with the action (\ref
{c4.4}) for $\mathcal{S}_{\mathrm{d}}$. If $\hbar \neq 0$, the expression
for the action (\ref{b3.1c}) is not the well defined, and dynamic equations
for $\mathcal{S}_{\mathrm{st}}$ are absent.

Dynamic equation for $\mathbf{u}$ is obtained from the action functional (%
\ref{c4.5}) by means of variation with respect to $\mathbf{u}$. If the
quantum constant $\hbar =0$, it follows from the dynamic equation for $%
\mathbf{u}$, that $\mathbf{u}=0$, and the action (\ref{c4.5}) reduces to the
form (\ref{c4.4}). In the general case $\hbar \neq 0$ we are to go to
independent variables $\mathbf{x}$\textbf{,} because $\mathbf{u}$ is a
function of $t,\mathbf{x}$. We obtain instead of (\ref{c4.5}) 
\begin{equation}
\mathcal{E}_{\mathrm{st}}\left[ \mathcal{S}_{\mathrm{st}}\right] :\qquad 
\mathcal{A}_{\mathcal{E}_{\mathrm{st}}\left[ \mathcal{S}_{\mathrm{st}}\right]
}\left[ \mathbf{x,u}\right] =\int \left\{ \frac{m}{2}\left( \frac{d\mathbf{x}%
}{dt}\right) ^{2}+\frac{m}{2}\mathbf{u}^{2}-\frac{\hbar }{2}\mathbf{\nabla u}%
\right\} \rho dtd\mathbf{x}  \label{c4.6}
\end{equation}
\textbf{\ } 
\begin{equation}
\rho =\frac{\partial \left( \xi _{1},\xi _{2},\xi _{3}\right) }{\partial
\left( x^{1},x^{2},x^{3}\right) }=\left( \frac{\partial \left(
x^{1},x^{2},x^{3}\right) }{\partial \left( \xi _{1},\xi _{2},\xi _{3}\right) 
}\right) ^{-1}  \label{c4.7}
\end{equation}
Variation of (\ref{c4.7}) with respect to $\mathbf{u}$ gives 
\begin{equation}
\frac{\delta \mathcal{A}_{\mathcal{E}_{\mathrm{st}}\left[ \mathcal{S}_{%
\mathrm{st}}\right] }}{\delta \mathbf{u}}=m\rho \mathbf{u+}\frac{\hbar }{2}%
\mathbf{\nabla }\rho =0  \label{c4.8}
\end{equation}

Resolving dynamic equation (\ref{c4.8}) with respect to $\mathbf{u}$ in the
form 
\begin{equation}
\mathbf{u=-}\frac{\hbar }{2m}\mathbf{\nabla }\ln \rho   \label{c4.9}
\end{equation}
we can eliminate the mean stochastic velocity $\mathbf{u}$ from the action (%
\ref{c4.6}). We obtain instead of (\ref{c4.6}) 
\begin{equation}
\mathcal{E}_{\mathrm{st}}\left[ \mathcal{S}_{\mathrm{st}}\right] :\qquad 
\mathcal{A}_{\mathcal{E}_{\mathrm{st}}\left[ \mathcal{S}_{\mathrm{st}}\right]
}\left[ \mathbf{x}\right] =\int \left\{ \frac{m}{2}\left( \frac{d\mathbf{x}}{%
dt}\right) ^{2}-U\left( \rho \mathbf{,\nabla }\rho \right) \right\} \rho dtd%
\mathbf{x}  \label{c4.10}
\end{equation}
where 
\begin{equation}
\rho U\left( \rho \mathbf{,\nabla }\rho \right) =-\frac{\hbar ^{2}}{8m}\frac{%
\left( \mathbf{\nabla }\rho \right) ^{2}}{\rho }-\frac{\hbar ^{2}}{4m}\rho 
\mathbf{\nabla }^{2}\ln \rho   \label{c4.11}
\end{equation}
and $\rho $ is defined by (\ref{c4.7}). Eliminating divergence, we obtain
instead of (\ref{c4.11}) 
\begin{equation}
U\left( \rho \mathbf{,\nabla }\rho \right) =\frac{\hbar ^{2}}{8m}\frac{%
\left( \mathbf{\nabla }\rho \right) ^{2}}{\rho ^{2}}+\frac{\hbar ^{2}}{4m}%
\frac{1}{\rho }\mathbf{\nabla }^{2}\rho   \label{c4.12}
\end{equation}
The last term in (\ref{c4.12}) does not give a contribution into dynamic
equations, and it may be omitted. The action (\ref{c4.10}) turns into 
\begin{equation}
\mathcal{E}_{\mathrm{st}}\left[ \mathcal{S}_{\mathrm{st}}\right] :\qquad 
\mathcal{A}_{\mathcal{E}_{\mathrm{st}}\left[ \mathcal{S}_{\mathrm{st}}\right]
}\left[ \mathbf{\xi }\right] =\int \left\{ \frac{m}{2}\left( \frac{d\mathbf{x%
}}{dt}\right) ^{2}-\frac{\hbar ^{2}}{8m}\frac{\left( \mathbf{\nabla }\rho
\right) ^{2}}{\rho ^{2}}\right\} \rho dtd\mathbf{x}  \label{c4.14}
\end{equation}
where variables $t,\mathbf{x}$ are independent variables, and variables $%
\mathbf{\xi }$ are considered to be dependent variables. The quantities $%
\rho $ and $\frac{d\mathbf{x}}{dt}$ are functions of the dependent variables 
$\mathbf{\xi }$ derivatives with respect to $t$ and $\mathbf{x}$%
\begin{equation}
\rho =\frac{\partial \left( \xi _{1},\xi _{2},\xi _{3}\right) }{\partial
\left( x^{1},x^{2},x^{3}\right) }  \label{c4.15}
\end{equation}
\begin{equation}
\frac{dx^{\alpha }}{dt}=\frac{\partial \left( x^{\alpha },\xi _{1},\xi
_{2},\xi _{3}\right) }{\partial \left( t,\xi _{1},\xi _{2},\xi _{3}\right) }%
=\rho ^{-1}\frac{\partial \left( x^{\alpha },\xi _{1},\xi _{2},\xi
_{3}\right) }{\partial \left( t,x^{1},x^{2},x^{3}\right) },\qquad \alpha
=1,2,3  \label{c4.16}
\end{equation}

Dynamic equations, generated by the action (\ref{c4.14}), are rather
complicated. However, in terms of the wave function the action (\ref{c4.14})
takes a more simple form \cite{R99}.

In terms of the two-component wave function $\psi $ 
\begin{equation}
\psi =\left\{ \psi _{1},\psi _{2}\right\} ,\qquad \psi ^{\ast }=\left\{ 
\begin{array}{c}
\psi _{1}^{\ast } \\ 
\psi _{2}^{\ast }
\end{array}
\right\} ,\qquad \rho \equiv \psi ^{\ast }\psi \equiv \psi _{1}^{\ast }\psi
_{1}+\psi _{2}^{\ast }\psi _{2},  \label{c4.17}
\end{equation}
the action (\ref{c4.16}) takes the form 
\begin{eqnarray}
\mathcal{E}_{\mathrm{st}}\left[ \mathcal{S}_{\mathrm{st}}\right] &:&\qquad 
\mathcal{A}_{\mathcal{E}_{\mathrm{st}}\left[ \mathcal{S}_{\mathrm{st}}\right]
}\left[ \psi ,\psi ^{\ast }\right] =\int \left\{ \frac{ib_{0}}{2}\left( \psi
^{\ast }\partial _{0}\psi -\partial _{0}\psi ^{\ast }\cdot \psi \right) -%
\frac{b_{0}^{2}}{2m}\mathbf{\nabla }\psi ^{\ast }\mathbf{\nabla }\psi \right.
\nonumber \\
&&\left. +\frac{b_{0}^{2}}{8m}\sum\limits_{\alpha =1}^{\alpha =3}(\mathbf{%
\nabla }s_{\alpha })^{2}\rho +\frac{b_{0}^{2}-\hbar ^{2}}{8\rho m}(\mathbf{%
\nabla }\rho )^{2}\right\} dtd\mathbf{x},  \label{c4.18}
\end{eqnarray}
where 
\begin{equation}
s_{\alpha }\equiv \frac{\psi ^{\ast }\sigma _{\alpha }\psi }{\rho },\qquad
\alpha =1,2,3,  \label{c4.19}
\end{equation}
and $\sigma _{\alpha }$ are the Pauli matrices 
\begin{equation}
\sigma _{1}=\left( 
\begin{array}{cc}
0 & 1 \\ 
1 & 0
\end{array}
\right) ,\qquad \sigma _{2}=\left( 
\begin{array}{cc}
0 & -i \\ 
i & 0
\end{array}
\right) ,\qquad \sigma _{3}=\left( 
\begin{array}{cc}
1 & 0 \\ 
0 & -1
\end{array}
\right)  \label{c4.20}
\end{equation}
Here the constant $b_{0}$ is an arbitrary constant. We transit from the
action (\ref{c4.14}) to the action (\ref{c4.18}) by means of the change of
variables, accompanied by the integration of dynamic equations and by the
appearance of three arbitrary functions $\mathbf{g}\left( \mathbf{\xi }%
\right) =\left\{ g^{1}\left( \mathbf{\xi }\right) ,g^{2}\left( \mathbf{\xi }%
\right) ,g^{3}\left( \mathbf{\xi }\right) \right\} $.

The change of variables, connecting dependent variables $\mathbf{\xi }$ and $%
\psi $, has the form (see Appendix A or \cite{R99}) 
\begin{equation}
\psi _{\alpha }=\sqrt{\rho }e^{i\varphi }u_{\alpha }(\mathbf{\xi }),\qquad
\psi _{\alpha }^{\ast }=\sqrt{\rho }e^{-i\varphi }u_{\alpha }^{\ast }(%
\mathbf{\xi }),\qquad \alpha =1,2,\ldots ,n,  \label{s1.1}
\end{equation}
\begin{equation}
\psi ^{\ast }\psi \equiv \sum_{\alpha =1}^{n}\psi _{\alpha }^{\ast }\psi
_{\alpha },  \label{s1.2}
\end{equation}
where (*) means the complex conjugate. The quantities $u_{\alpha }(\mathbf{%
\xi })$, $\alpha =1,2,...n$ are functions of only variables $\mathbf{\xi }$,
and satisfy the relations 
\begin{equation}
-{\frac{i}{2}}\sum_{\alpha =1}^{n}\left( u_{\alpha }^{\ast }\frac{\partial
u_{\alpha }}{\partial \xi _{\beta }}-\frac{\partial u_{\alpha }^{\ast }}{%
\partial \xi _{\beta }}u_{\alpha }\right) =g^{\beta }(\mathbf{\xi }),\qquad
\beta =1,2,3,\qquad \sum_{\alpha =1}^{n}u_{\alpha }^{\ast }u_{\alpha }=1.
\label{s5.5}
\end{equation}
Here $\varphi $ is the new dependent variable, appearing from the fictitious
temporal Lagrangian coordinate $\xi _{0}$, and $b_{0}$ is an arbitrary
constant. The number $n$ is such a natural number that equations (\ref{s5.5}%
) admit a solution. In general, $n$ depends on the form of the arbitrary
integration functions $\mathbf{g}=\{g^{\beta }(\mathbf{\xi })\}$, $\beta
=1,2,3$.

The meaning of the wave function $\psi $ is not clear, and interpretation is
produced on the basis of the action (\ref{c4.6}) or (\ref{c4.14}), where
meaning of all quantities is quite clear. The action (\ref{c4.6}) describes
the flow of some fluid with the density $\rho $, determined by the relation (%
\ref{c4.7}), and the flux density 
\begin{equation}
\mathbf{j}=\rho \frac{d\mathbf{x}}{dt}  \label{c4.21}
\end{equation}

In terms of the wave function $\psi $ these quantities have the form 
\begin{equation}
\rho =\psi ^{\ast }\psi ,\qquad \mathbf{j=-}\frac{ib_{0}}{2m}\left( \psi
^{\ast }\mathbf{\nabla }\psi -\mathbf{\nabla }\psi ^{\ast }\cdot \psi \right)
\label{c4.22}
\end{equation}

The functions $\mathbf{g}$ determine vorticity of the fluid flow. If $%
\mathbf{g}=0$, equations (\ref{s5.5}) have the solution $u_{1}=1$, $%
u_{\alpha }=0$, \ $\alpha =2,3,...n$. In this case the function $\psi $ may
have one component (other components vanish), and the fluid flow is
irrotational. The function $\psi $ has the form 
\begin{equation}
\psi =\sqrt{\rho }e^{i\varphi },\qquad \psi ^{\ast }=\sqrt{\rho }%
e^{-i\varphi }  \label{c4.23}
\end{equation}
and the fluid velocity 
\begin{equation}
\mathbf{v}=\frac{\mathbf{j}}{\rho }=\mathbf{\nabla }\frac{b_{0}\varphi }{m}
\label{c4.24}
\end{equation}
has the potential $b_{0}\varphi /m$.

In the partial case of the irrotational fluid flow 
\begin{equation}
s_{\alpha }\equiv \frac{\psi ^{\ast }\sigma _{\alpha }\psi }{\rho }=\text{%
const},\qquad \alpha =1,2,3  \label{c4.25}
\end{equation}
and the action (\ref{c4.18}) turns into the action 
\begin{equation}
\mathcal{A}_{\mathcal{S}_{\mathrm{q}}}\left[ \psi ,\psi ^{\ast }\right]
=\int \left\{ \frac{ib_{0}}{2}\left( \psi ^{\ast }\partial _{0}\psi
-\partial _{0}\psi ^{\ast }\cdot \psi \right) -\frac{b_{0}^{2}}{2m}\mathbf{%
\nabla }\psi ^{\ast }\mathbf{\nabla }\psi +\frac{b_{0}^{2}-\hbar ^{2}}{8\rho
m}(\mathbf{\nabla }\rho )^{2}\right\} dtd\mathbf{x},  \label{c4.26}
\end{equation}
If we choose the arbitrary constant $b_{0}$ in the form $b_{0}=\hbar $, the
action (\ref{c4.26}) turns into the action 
\begin{equation}
\mathcal{A}_{\mathcal{S}_{\mathrm{q}}}\left[ \psi ,\psi ^{\ast }\right]
=\int \left\{ \frac{i\hbar }{2}\left( \psi ^{\ast }\partial _{0}\psi
-\partial _{0}\psi ^{\ast }\cdot \psi \right) -\frac{\hbar ^{2}}{2m}\mathbf{%
\nabla }\psi ^{\ast }\mathbf{\nabla }\psi \right\} dtd\mathbf{x},
\label{c4.27}
\end{equation}
having the Schr\"{o}dinger equation 
\begin{equation}
i\hbar \partial _{0}\psi =-\frac{\hbar ^{2}}{2m}\mathbf{\nabla }^{2}\psi
\label{c4.28}
\end{equation}
as the dynamic equation. Expressions (\ref{c4.22}) for the density and the
particle flux turn into the conventional expressions 
\begin{equation}
\rho =\psi ^{\ast }\psi ,\qquad \mathbf{j=-}\frac{i\hbar }{2m}\left( \psi
^{\ast }\mathbf{\nabla }\psi -\mathbf{\nabla }\psi ^{\ast }\cdot \psi \right)
\label{c4.29}
\end{equation}

Interpretation of all quantities is obtained on the basis the fact, that the
quantum description in terms of the Schr\"{o}dinger equation is the special
case of the statistical description in terms of the statistical ensemble (%
\ref{c4.5}).

Can we obtain the statistical ensemble (\ref{c4.5}) from the statistical
ensemble (\ref{c4.4}) by means of the change $m\rightarrow m_{\mathrm{eff}}$%
? It is possible, if we represent the nonrelativistic action (\ref{c4.4}) as
the nonrelativistic approximation 
\begin{equation}
\mathcal{E}_{\mathrm{d}}\left[ \mathcal{S}_{\mathrm{d}}\right] :\qquad 
\mathcal{A}_{\mathcal{E}_{\mathrm{d}}\left[ \mathcal{S}_{\mathrm{d}}\right] }%
\left[ \mathbf{x}\right] =\int \left\{ -mc^{2}+\frac{m}{2}\left( \frac{d%
\mathbf{x}}{dt}\right) ^{2}\right\} dtd\mathbf{\xi }  \label{c4.30}
\end{equation}
of the relativistic action 
\begin{equation}
\mathcal{E}_{\mathrm{d}}\left[ \mathcal{S}_{\mathrm{d}}\right] :\qquad 
\mathcal{A}_{\mathcal{E}_{\mathrm{d}}\left[ \mathcal{S}_{\mathrm{d}}\right] }%
\left[ \mathbf{x}\right] =-\int mc^{2}\sqrt{1-\frac{1}{c^{2}}\left( \frac{d%
\mathbf{x}}{dt}\right) ^{2}}dtd\mathbf{\xi }  \label{c4.31}
\end{equation}
The action (\ref{c4.5}) is obtained from the action (\ref{c4.30}) as a
result of the change 
\begin{equation}
m\rightarrow m_{\mathrm{eff}}=m\left( 1-\frac{\mathbf{u}^{2}}{2c^{2}}+\frac{%
\hbar }{2mc^{2}}\mathbf{\nabla u}\right)  \label{c4.32}
\end{equation}
Practically, the change is produced only in the first term of the action (%
\ref{c4.30}), because the change in the second term gives additional term of
the order of $c^{-2}$, which is small in the nonrelativistic approximation.
Another version of the change in the action (\ref{c4.30}) has the form 
\begin{equation}
m\rightarrow m_{\mathrm{eff}}=m\left( 1-\frac{\hbar ^{2}}{8m^{2}c^{2}}\frac{%
\left( \mathbf{\nabla }\rho \right) ^{2}}{\rho ^{2}}-\frac{\hbar ^{2}}{%
4m^{2}c^{2}}\frac{1}{\rho }\mathbf{\nabla }^{2}\rho \right)  \label{c4.33}
\end{equation}
or 
\begin{equation}
m\rightarrow m_{\mathrm{eff}}=m\left( 1+\frac{\hbar ^{2}}{8m^{2}c^{2}}\frac{%
\left( \mathbf{\nabla }\rho \right) ^{2}}{\rho ^{2}}\right)  \label{c4.34}
\end{equation}
The relation (\ref{c4.33}) is obtained from (\ref{c4.32}) after substitution
of (\ref{c4.9}). Producing the change (\ref{c4.34}) in the action (\ref
{c4.30}), we obtain in the nonrelativistic approximation the action (\ref
{c4.14}).

In the relativistic case instead of the change (\ref{c4.32}) we have 
\begin{equation}
m^{2}\rightarrow m_{\mathrm{eff}}^{2}=m^{2}\left( 1+u_{l}u^{l}+\lambda
\partial _{l}u^{l}\right) ,\qquad \lambda =\frac{\hbar }{mc}  \label{c4.35}
\end{equation}
where the variables $u^{k}=u^{k}\left( x\right) =u^{k}\left( t,\mathbf{x}%
\right) $, $k=0,1,2,3$ are new dependent variables, describing the mean
value of the stochastic component of the particle 4-velocity. The change (%
\ref{c4.35}) in the action (\ref{c4.31}) for the statistical ensemble of
free relativistic particles leads finally to the action \cite{R003} 
\begin{equation}
\mathcal{A}\left[ \psi ,\psi ^{\ast }\right] =\int \left\{ \hbar
^{2}\partial _{k}\psi ^{\ast }\partial ^{k}\psi -m^{2}c^{2}\rho -\frac{\hbar
^{2}}{4}\left( \partial _{l}s_{\alpha }\right) \left( \partial ^{l}s_{\alpha
}\right) \rho \right\} d^{4}x  \label{c4.36}
\end{equation}
where $\psi $ is the two-component wave function (\ref{s1.1}) - (\ref{s5.5}%
). The variables $\rho $, $s_{\alpha }$ are defined by the relation (\ref
{c4.22}) and, besides, the constant $b_{0}=\hbar $. The action (\ref{c4.36})
is the action for the statistical ensemble of free stochastic relativistic
particles. In the case of irrotational flow, when the wave function $\psi $
may be one-component, $s_{\alpha }=$const, and the dynamic equation for the
action (\ref{c4.36}) is the Klein-Gordon equation. 
\begin{equation}
\hbar ^{2}\partial _{k}\partial ^{k}\psi +m^{2}c^{2}\psi =0  \label{c4.37}
\end{equation}

\section{Determination of the effective mass}

We are going to show, that the change (\ref{c4.34}) follows from the form of
the world function (\ref{b3.3}). In reality in \cite{R91} the inverse
problem has been solved. What is the geometry of the uniform space-time, if
the statistical description of free nonrelativistic particles leads to the
quantum description in terms of the Schr\"{o}dinger equation? Having solved
this problem, we obtained the world function (\ref{b3.3}). Now we show that
the effective mass $m_{\mathrm{eff}}$ of the nonrelativistic particle is
determined by the relation (\ref{c4.34}).

Mathematical formalism of theoretical physics is suited for application in
the Minkowski space-time. Mathematical formalism for work in the distorted
space-time $V_{\mathrm{d}}$ with the world function (\ref{b3.3}) is absent
now. We are forced to work in the Minkowski space-time, using conventional
technique and taking into account distortion of the space-time by means of
some corrections.

Let introduce the notion of the adduced vector $\vec{p}=\vec{p}\left(
a,P_{0},P_{1}\right) $ as a totality of a real or imaginary number $a$ and
two points $\left\{ P_{0},P_{1}\right\} $ 
\begin{equation}
\vec{p}=\vec{p}\left( a,P_{0},P_{1}\right) =a\left( \mathbf{P}_{0}\mathbf{P}%
_{1}\right) =a\mathbf{P}_{0}\mathbf{P}_{1}  \label{c5.2}
\end{equation}
The number $a$ is called the gauge of the adduced vector. The vector $%
\mathbf{P}_{0}\mathbf{P}_{1}$ is a partial case of the adduced vector $a%
\mathbf{P}_{0}\mathbf{P}_{1}$ with the gauge $a=1$. The scalar product of
two adduced vectors $a_{1}\mathbf{P}_{0}\mathbf{P}_{1}$ and $a_{2}\mathbf{Q}%
_{0}\mathbf{Q}_{1}$ is defined by the relation 
\begin{equation}
\left( a_{1}\mathbf{P}_{0}\mathbf{P}_{1}.a_{2}\mathbf{Q}_{0}\mathbf{Q}%
_{1}\right) =a_{1}a_{2}\left( \mathbf{P}_{0}\mathbf{P}_{1}.\mathbf{Q}_{0}%
\mathbf{Q}_{1}\right)  \label{c5.3}
\end{equation}

We shall consider statistical ensemble of relativistic particles, described
by the action (\ref{c4.31}) with the oriented mass $m_{\mathrm{o}}$, defined
by the relation 
\begin{equation}
m_{\mathrm{o}}=b\mu _{\mathrm{o}},\qquad \mu _{\mathrm{o}}=\left( \mathbf{P}%
_{0}\mathbf{P}_{1}.\vec{u}\left( R\right) \right)  \label{c5.1}
\end{equation}
where $\vec{u}=\vec{u}\left( R\right) $ is the unit adduced vector of the
4-velocity at the point $R\in \mathcal{T}_{\left[ P_{0}P_{1}\right] }$, $%
\mathbf{P}_{0}\mathbf{P}_{1}$ is the momentum vector, divided by the speed
of the light $c$. The quantity $m_{\mathrm{o}}$ is called the oriented mass
because it depends on the mutual orientation of the momentum vector and of
the 4-velocity. The oriented mass $m_{\mathrm{o}}$ has different sign for
the particle and for the antiparticle.

The 4-velocity $\vec{u}=\vec{u}\left( R\right) $ is the unit adduced vector
inside the segment $\mathcal{T}_{\left[ P_{0}P_{1}\right] }$ in the
space-time $V_{\mathrm{d}}$%
\begin{equation}
\vec{u}\left( R\right) =\left| \mathbf{P}_{0}\mathbf{R}\right| _{\mathrm{d}%
}^{-1}\mathbf{P}_{0}\mathbf{R}=\left( \mathbf{P}_{0}\mathbf{R}.\mathbf{P}_{0}%
\mathbf{R}\right) _{\mathrm{d}}^{-1/2}\mathbf{P}_{0}\mathbf{R,\qquad }R\in 
\mathcal{T}_{\left[ P_{0}P_{1}\right] }  \label{c5.4}
\end{equation}
\begin{equation}
\left( \vec{u}\left( R\right) .\vec{u}\left( R\right) \right) _{\mathrm{d}}=1
\label{c5.5}
\end{equation}

The particle mass, defined by the relation (\ref{c5.1}), is different in $V_{%
\mathrm{d}}$ and in $V_{\mathrm{M}}$. As far as $R\in \mathcal{T}_{\left[
P_{0}P_{1}\right] }$ and, hence, $\mathbf{P}_{0}\mathbf{R\uparrow \uparrow }%
_{\mathrm{d}}\mathbf{P}_{0}\mathbf{P}_{1}$. 
\begin{equation}
\left( \mathbf{P}_{0}\mathbf{P}_{1}.\mathbf{P}_{0}\mathbf{R}\right) _{%
\mathrm{d}}=\left| \mathbf{P}_{0}\mathbf{P}_{1}\right| _{\mathrm{d}}\cdot
\left| \mathbf{P}_{0}\mathbf{R}\right| _{\mathrm{d}},  \label{c5.6}
\end{equation}
we obtain for $m_{\mathrm{od}}$%
\begin{eqnarray}
m_{\mathrm{od}} &=&b\left( \mathbf{P}_{0}\mathbf{P}_{1}.\vec{u}\left(
R\right) \right) _{\mathrm{d}}=b\left| \mathbf{P}_{0}\mathbf{R}\right| _{%
\mathrm{d}}^{-1}\left( \mathbf{P}_{0}\mathbf{P}_{1}.\mathbf{P}_{0}\mathbf{R}%
\right) _{\mathrm{d}}  \nonumber \\
&=&b\left| \mathbf{P}_{0}\mathbf{R}\right| _{\mathrm{d}}^{-1}\cdot \left| 
\mathbf{P}_{0}\mathbf{P}_{1}\right| _{\mathrm{d}}\cdot \left| \mathbf{P}_{0}%
\mathbf{R}\right| _{\mathrm{d}}=b\left| \mathbf{P}_{0}\mathbf{P}_{1}\right|
_{\mathrm{d}}=b\mu _{\mathrm{d}}  \label{c5.7}
\end{eqnarray}
where $b$ is the constant, defined by (\ref{b3.14}).

If the point $R$ on the segment $\mathcal{T}_{\left[ P_{0}P_{1}\right] }$ is
not close to the ends $P_{0}$ and $P_{1}$, (i.e. \ $\left| \mathbf{P}_{0}%
\mathbf{R}\right| ^2_{\mathrm{d}}>2\sigma _{0},\;\left| \mathbf{P}_{0}%
\mathbf{P}_{1}\right| ^2_{\mathrm{d}}>2\sigma _{0}$) and relation (\ref{b3.7}%
) is satisfied, we obtain for the oriented mass $m_{\mathrm{oM}}$ in $V_{%
\mathrm{M}}$%
\begin{eqnarray}
m_{\mathrm{oM}} &=&b\left( \mathbf{P}_{0}\mathbf{P}_{1}.\vec{u}\left(
R\right) \right) _{\mathrm{M}}=b\left| \mathbf{P}_{0}\mathbf{R}\right| _{%
\mathrm{d}}^{-1}\left( \mathbf{P}_{0}\mathbf{P}_{1}.\mathbf{P}_{0}\mathbf{R}%
\right) _{\mathrm{M}}  \nonumber \\
&=&b\left| \mathbf{P}_{0}\mathbf{R}\right| _{\mathrm{d}}^{-1}\left( \left( 
\mathbf{P}_{0}\mathbf{P}_{1}.\mathbf{P}_{0}\mathbf{R}\right) _{\mathrm{d}%
}-2d\right)  \nonumber \\
&=&b\left| \mathbf{P}_{0}\mathbf{P}_{1}\right| _{\mathrm{d}}-2db\left| 
\mathbf{P}_{0}\mathbf{R}\right| _{\mathrm{d}}^{-1}=b\mu _{\mathrm{d}}-\frac{%
2bd}{\left| \mathbf{P}_{0}\mathbf{R}\right| _{\mathrm{d}}}  \label{c5.8}
\end{eqnarray}

Thus, the particle mass $m_{\mathrm{oM}}$, defined by the relation (\ref
{c5.1}) and calculated in $V_{\mathrm{M}}$ depends on the point $R$ on the
surface of the segment $\mathcal{T}_{\left[ P_{0}P_{1}\right] }$. We use in
the action (\ref{c4.31}) some effective mass $m_{\mathrm{eff}}$, calculated
in accordance with (\ref{c5.8}) in the Minkowski space-time $V_{\mathrm{M}}$
by means of the relation 
\begin{equation}
m_{\mathrm{eff}}=b\left( \mathbf{P}_{0}\mathbf{P}_{1}.\vec{u}_{\mathrm{eff}%
}\right) _{\mathrm{M}}  \label{c5.9}
\end{equation}
where the adduced vector $\vec{u}_{\mathrm{eff}}$ is the mean 4-velocity
inside the segment $\mathcal{T}_{\left[ P_{0}P_{1}\right] }$.

Let the point $P$ be the center of the segment $\mathcal{T}_{\left[
P_{0}P_{1}\right] }$, as it shown in figure 1. The points $P^{\prime }$ and $%
P^{\prime \prime }$ are centers of segments $\mathcal{T}_{\left[
P_{0}^{\prime }P_{1}^{\prime }\right] }$, $\mathcal{T}_{\left[ P_{0}^{\prime
\prime }P_{1}^{\prime \prime }\right] }$, of adjacent world tubes of the
statistical ensemble. We consider nonrelativistic case, and the vectors $%
\mathbf{P}_{0}\mathbf{P}_{1}$, $\mathbf{P}_{0}^{\prime }\mathbf{P}%
_{1}^{\prime }$, $\mathbf{P}_{0}^{\prime \prime }\mathbf{P}_{1}^{\prime
\prime }$ may be considered to be parallel in $V_{\mathrm{M}}$. Let segments 
$\mathcal{T}_{\left[ P_{0}P_{1}\right] }$, $\mathcal{T}_{\left[
P_{0}^{\prime }P_{1}^{\prime }\right] }$, $\mathcal{T}_{\left[ P_{0}^{\prime
\prime }P_{1}^{\prime \prime }\right] }$ be placed in such a way, that $P\in 
\mathcal{T}_{\left[ P_{0}^{\prime }P_{1}^{\prime }\right] }$, $P\in \mathcal{%
T}_{\left[ P_{0}^{\prime \prime }P_{1}^{\prime \prime }\right] }$. The
4-velocity of the segment $\mathcal{T}_{\left[ P_{0}^{\prime }P_{1}^{\prime }%
\right] }$, determined by the vector $\mathbf{P}_{0}^{\prime }\mathbf{P}$,
and the 4-velocity of the segment $\mathcal{T}_{\left[ P_{0}^{\prime \prime
}P_{1}^{\prime \prime }\right] }$, determined by the vector $\mathbf{P}%
_{0}^{\prime \prime }\mathbf{P}$\textbf{,} make a contribution in the
effective 4-velocity $\vec{u}_{\mathrm{eff}}$ of the segment $\mathcal{T}_{%
\left[ P_{0}P_{1}\right] }$. We suppose that the origin of the effective
4-velocity vector $\vec{u}_{\mathrm{eff}}$ is placed at the point $P$. Let
the spatial distance between the points $P,P^{\prime }$ and $P,P^{\prime
\prime }$ be $l$. According to the relation (\ref{b3.7a}) we obtain 
\begin{equation}
l=\sqrt{-\left| \mathbf{PP}^{\prime }\right| ^{2}}=\sqrt{-\left| \mathbf{PP}%
^{\prime \prime }\right| ^{2}}=r\left( 0.5\right) =\sqrt{\frac{3d}{2}}=\sqrt{%
\frac{3\hbar }{4bc}}  \label{c5.10}
\end{equation}

We choose the coordinate system with the origin at the point $P$ and with
the time axis directed along the vector $\mathbf{P}_{0}\mathbf{P}_{1}$. In
the space-time $V_{\mathrm{M}}$ in this coordinate system we have covariant
components of $\mathbf{P}_{0}\mathbf{P}_{1}$ 
\begin{equation}
\left( \mathbf{P}_{0}\mathbf{P}_{1}\right) _{k}=\left\{ \mu _{\mathrm{d}%
}c,0,\right\}  \label{c5.10a}
\end{equation}
The contravariant coordinates of the 4-velocity of the segment $\mathcal{T}_{%
\left[ P_{0}^{\prime }P_{1}^{\prime }\right] }$ $\mathbf{x}$ have the form 
\begin{equation}
u^{k}=\left\{ u^{0},\mathbf{u}\right\} =\left\{ \frac{1}{c\sqrt{1-\frac{%
\mathbf{v}^{2}}{c^{2}}}},-\frac{\mathbf{v}}{c\sqrt{1-\frac{\mathbf{v}^{2}}{%
c^{2}}}}\right\} ,\qquad \mathbf{v=-}\frac{2\mathbf{x}}{\mu _{\mathrm{d}}}c
\label{c5.12}
\end{equation}
where $\mathbf{x}$ are the spatial coordinates of the point $P^{\prime }$.
The effective 4-velocity at the point $P$ is a sum of contributions of all
segments $\mathcal{T}_{\left[ P_{0}^{\prime }P_{1}^{\prime }\right] }$%
\begin{equation}
u_{\mathrm{eff}}^{0}=A\int \rho \left( \mathbf{x}\right) \delta \left( l^{2}-%
\mathbf{x}^{2}\right) u^{0}\left( \mathbf{x}\right) d\mathbf{x,}
\label{c5.14}
\end{equation}
\begin{equation}
l=\sqrt{\frac{3\hbar }{4bc}}  \label{c5.14a}
\end{equation}
\begin{equation}
\mathbf{u}_{\mathrm{eff}}=A\int \rho \left( \mathbf{x}\right) \delta \left(
l^{2}-\mathbf{x}^{2}\right) \mathbf{u}\left( \mathbf{x}\right) d\mathbf{x}%
=-A\int \frac{\rho \left( \mathbf{x}\right) \delta \left( l^{2}-\mathbf{x}%
^{2}\right) }{\sqrt{1-\left( \frac{2\mathbf{x}}{\mu _{\mathrm{d}}}\right)
^{2}}}\frac{2\mathbf{x}}{\mu _{\mathrm{d}}}d\mathbf{x}  \label{c5.15}
\end{equation}
where the quantity $\rho $ is the density of world lines in the statistical
ensemble, and the quantity $A$ is determined from the condition of
normalization of $\vec{u}_{\mathrm{eff}}\left( P\right) $ 
\begin{equation}
c^{2}\left( u_{\mathrm{eff}}^{0}\right) ^{2}-\mathbf{u}_{\mathrm{eff}}^{2}=1
\label{c5.16}
\end{equation}

Supposing that $\rho \left( \mathbf{x}\right) $ changes slowly and expanding 
$\rho \left( \mathbf{x}\right) $ in a series over $\mathbf{x}$, we obtain
from (\ref{c5.14}), (\ref{c5.15}) 
\begin{equation}
u_{\mathrm{eff}}^{0}=A\int \frac{\rho }{c\sqrt{1-\left( \frac{2\mathbf{x}}{%
\mu _{\mathrm{d}}}\right) ^{2}}}\delta \left( l^{2}-\mathbf{x}^{2}\right) d%
\mathbf{x}=\frac{2\pi A\rho l}{c\sqrt{1-\left( \frac{2l}{\mu _{\mathrm{d}}}%
\right) ^{2}}}  \label{c5.17}
\end{equation}
\begin{equation}
\mathbf{u}_{\mathrm{eff}}=-A\int \frac{\left( \mathbf{x\nabla }\right) \rho 
}{\sqrt{1-\left( \frac{2\mathbf{x}}{\mu _{\mathrm{d}}}\right) ^{2}}}\delta
\left( l^{2}-\mathbf{x}^{2}\right) \frac{2\mathbf{x}}{\mu _{\mathrm{d}}}cd%
\mathbf{x}=-\frac{4\pi Al^{3}\mathbf{\nabla }\rho }{3\mu _{\mathrm{d}}\sqrt{%
1-\left( \frac{2l}{\mu _{\mathrm{d}}}\right) ^{2}}}  \label{c5.18}
\end{equation}
where $\rho $ is the value of the density at the point $P$. Substituting (%
\ref{c5.17}), (\ref{c5.18}) in (\ref{c5.16}) and using (\ref{c5.14a}), we
obtain 
\begin{equation}
A=\frac{\sqrt{1-\left( \frac{2l}{\mu _{\mathrm{d}}}\right) ^{2}}}{2\pi l\rho 
\sqrt{1-\left( \frac{\hbar }{2m_{\mathrm{d}}c}\mathbf{\nabla }\ln \rho
\right) ^{2}}}  \label{c5.19}
\end{equation}
Substituting (\ref{c5.19}) in (\ref{c5.17}), we obtain 
\begin{equation}
u_{\mathrm{eff}}^{0}=\frac{1}{c\sqrt{1-\left( \frac{\hbar }{2m_{\mathrm{d}}c}%
\mathbf{\nabla }\ln \rho \right) ^{2}}}  \label{c5.20}
\end{equation}

It follows from (\ref{c5.9}), (\ref{c5.10a}) and (\ref{c5.20}) 
\begin{equation}
m_{\mathrm{eff}}=m_{\mathrm{d}}cu_{\mathrm{eff}}^{0}=m_{\mathrm{d}}\frac{1}{%
\sqrt{1-\left( \frac{\hbar }{2m_{\mathrm{d}}c}\mathbf{\nabla }\ln \rho
\right) ^{2}}}=m_{\mathrm{d}}\left( 1+\frac{\hbar ^{2}}{8m_{\mathrm{d}%
}^{2}c^{2}}\left( \mathbf{\nabla }\ln \rho \right) ^{2}\right)  \label{c5.21}
\end{equation}
This result coincides with the relation (\ref{c4.34}).

We admit that there are another methods of calculation of the value of $m_{%
\mathrm{eff}}$, which give another result. In this case we should choose
another world function of the space-time $V_{\mathrm{d}}$, which leads to
the result (\ref{c5.21}), because we know that the effective mass,
determined by the relation (\ref{c5.21}) agrees with the experimental data.
We know about the distorted space-time geometry only that it generates
stochastic motion of free particles. Information on its world function is
obtained from the demand that the world function leads to the effective
mass, which is determined by the relation (\ref{c5.21}).

Further development of the statistical description of geometrical
stochasticity leads to a creation of the model conception of quantum
phenomena (MCQP), which relates to the conventional quantum theory
approximately in the same way as the statistical physics relates to the
axiomatic thermodynamics. MCQP is the well defined relativistic conception
with effective methods of investigation \cite{R03}, whereas the conventional
quantum theory is not well defined, because it uses incorrect space-time
geometry, whose incorrectness is compensated by additional hypotheses
(quantum principles). Besides, it has problems with application of the
nonrelativistic quantum mechanics technique to the description of
relativistic phenomena.

The geometry $\mathcal{G}_{\mathrm{d}}$ is a homogeneous geometry as well as
the Minkowski geometry, because the world function $\sigma _{\mathrm{d}}$ is
invariant with respect to all coordinate transformations, with respect to
which the world function $\sigma _{\mathrm{M}}$ is invariant. In this
connection the question arises, whether one could invent some axiomatics for 
$\mathcal{G}_{\mathrm{d}}$ and derive the geometry $\mathcal{G}_{\mathrm{d}}$
from this axiomatics by means of proper reasonings. Note that such an
axiomatics is to depend on the parameter $d$, because the world function $%
\sigma _{\mathrm{d}}$ depends on this parameter. If $d=0$, this axiomatics
is to coincide with the axiomatics of the Minkowski geometry $\mathcal{G}_{%
\mathrm{M}}$. If $d\neq 0$, this axiomatics cannot coincide with the
axiomatics of $\mathcal{G}_{\mathrm{M}}$, because some axioms of $\mathcal{G}%
_{\mathrm{M}}$ are not satisfied in this case. In general, the invention of
axiomatics, depending on the parameter $d$ and in the general case on the
distortion function $D$, seems to be a very difficult problem. Besides, why
invent the axiomatics? We had derived the axiomatics for the proper
Euclidean geometry, when we constructed it before. There is no necessity to
repeat this process any time, when we construct a new geometry. It is
sufficient to apply the deformation principle to the constructed Euclidean
geometry written $\sigma $-immanently. Application of the deformation
principle to the Euclidean geometry is a very simple and general procedure,
which is not restricted by continuity, convexity and other artificial
constraints, generated by our preconceived approach to the physical
geometry. (Bias of the approach is displayed in the antecedent supposition
on the one-dimensionality of any straight line in any physical geometry,
which reminds the statement of the ancient Egyptians that all rivers flow
towards the North).

Thus, we have seen that the nondegeneracy of the physical geometry as well
as non-one-dimensionality of the straight line are properties of the real
physical geometries. The proper Euclidean geometry is a ground for all
physical geometries. Although it is a degenerate geometry, it is beyond
reason to deny an existence of nondegenerate physical geometries.

Thus, the deformation principle together with the $\sigma $-immanent
description appears to be a very effective mathematical tool for
construction of physical geometries.

\begin{enumerate}
\item  The deformation principle uses results obtained at construction of
the proper Euclidean geometry and does not add any additional supposition on
properties of geometrical objects.

\item  The deformation principle uses only the real characteristic of the
physical geometry -- its world function and does not use any additional
means of description.

\item  The deformation principle is very simple and allows one to
investigate only that part of geometry which one is interested in.

\item  Application of the deformation principle allows one to obtain the
true space-time geometry, whose unexpected properties cannot be obtained at
the conventional approach to physical geometry.
\end{enumerate}

\appendix
\renewcommand{\theequation}{\Alph{section}.\arabic{equation}} %
\renewcommand{\thesection}{Appendix \Alph{section}.}

\section{Transformation of the action for the statistical ensemble.}

To transform the action (\ref{c4.14}) to the description in terms of the
wave function, we rewrite it in the form 
\begin{equation}
\mathcal{E}_{\mathrm{st}}\left[ \mathcal{S}_{\mathrm{st}}\right] :\qquad 
\mathcal{A}_{\mathcal{E}_{\mathrm{st}}\left[ \mathcal{S}_{\mathrm{st}}\right]
}\left[ \mathbf{x}\right] =\int \left\{ \frac{m}{2}\left( \frac{d\mathbf{x}}{%
dt}\right) ^{2}-\frac{\hbar ^{2}}{8m}\frac{\left( \mathbf{\nabla }\rho
\right) ^{2}}{\rho ^{2}}\right\} dtd\mathbf{\xi }  \label{A.01}
\end{equation}
where 
\begin{equation}
\rho =\frac{\partial \left( \xi _{1},\xi _{2},\xi _{3}\right) }{\partial
\left( x^{1},x^{2},x^{3}\right) }=\left( \frac{\partial \left(
x^{1},x^{2},x^{3}\right) }{\partial \left( \xi _{1},\xi _{2},\xi _{3}\right) 
}\right) ^{-1}  \label{A.02}
\end{equation}
We introduce the independent variable $\xi _{0}$ instead of the variable $%
t=x^{0}$ and rewrite the action (\ref{A.01}) in the form 
\begin{equation}
\mathcal{A}_{\mathcal{E}_{\mathrm{st}}\left[ \mathcal{S}_{\mathrm{st}}\right]
}\left[ x\right] =\int \left\{ \frac{m\dot{x}^{\alpha }\dot{x}^{\alpha }}{2%
\dot{x}^{0}}-\frac{\hbar ^{2}}{8m}\frac{\left( \mathbf{\nabla }\rho \right)
^{2}}{\rho ^{2}}\right\} d^{4}\xi ,\qquad \dot{x}^{k}\equiv \frac{\partial
x^{k}}{\partial \xi _{0}}  \label{A.1}
\end{equation}
where $\xi =\left\{ \xi _{0},\mathbf{\xi }\right\} =\left\{ \xi _{k}\right\} 
$,\ \ $k=0,1,2,3$, $x=\left\{ x^{k}\left( \xi \right) \right\} $,\ \ $%
k=0,1,2,3$. Here and in what follows, a summation over repeated Greek
indices is produced $(1-3)$.

Let us consider variables $\xi =\xi \left( x\right) $ in (\ref{A.1}) as
dependent variables and variables $x$ as independent variables. Let the
Jacobian 
\begin{equation}
J=\frac{\partial \left( \xi _{0},\xi _{1},\xi _{2},\xi _{3}\right) }{%
\partial \left( x^{0},x^{1},x^{2},x^{3}\right) }=\det \left| \left| \xi
_{i,k}\right| \right| ,\qquad \xi _{i,k}\equiv \partial _{k}\xi _{i},\qquad
i,k=0,1,2,3  \label{A.3}
\end{equation}
be considered to be a multilinear function of $\xi _{i,k}$. Then 
\begin{equation}
d^{4}\xi =Jd^{4}x,\qquad \dot{x}^{i}\equiv \frac{\partial x^{i}}{\partial
\xi _{0}}\equiv \frac{\partial \left( x^{i},\xi _{1},\xi _{2},\xi
_{3}\right) }{\partial \left( \xi _{0},\xi _{1},\xi _{2},\xi _{3}\right) }%
=J^{-1}\frac{\partial J}{\partial \xi _{0,i}},\qquad i=0,1,2,3  \label{A.4}
\end{equation}
After transformation to dependent variables $\xi $ the action (\ref{A.1})
takes the form 
\begin{equation}
\mathcal{A}_{\mathcal{E}_{\mathrm{st}}\left[ \mathcal{S}_{\mathrm{st}}\right]
}\left[ \xi \right] =\int \left\{ \frac{m}{2}\frac{\partial J}{\partial \xi
_{0,\alpha }}\frac{\partial J}{\partial \xi _{0,\alpha }}\left( \frac{%
\partial J}{\partial \xi _{0,0}}\right) ^{-1}-\frac{\hbar ^{2}}{8m}\frac{%
\left( \mathbf{\nabla }\rho \right) ^{2}}{\rho }\right\} d^{4}x\mathbf{,}
\label{A.5}
\end{equation}
Here the dependent variable $\xi _{0}$ is fictitious

We introduce new variables 
\begin{equation}
j^{k}=\frac{\partial J}{\partial \xi _{0,k}},\qquad k=0,1,2,3,\qquad \rho
=j^{0}  \label{A.6}
\end{equation}
by means of Lagrange multipliers $p_{k}$%
\begin{equation}
\mathcal{A}_{\mathcal{E}_{\mathrm{st}}\left[ \mathcal{S}_{\mathrm{st}}\right]
}\left[ \xi ,j,p\right] =\int \left\{ \frac{m}{2}\frac{j^{\alpha }j^{\alpha }%
}{j^{0}}-\frac{\hbar ^{2}}{8m}\frac{\left( \mathbf{\nabla }\rho \right) ^{2}%
}{\rho }+p_{k}\left( \frac{\partial J}{\partial \xi _{0,k}}-j^{k}\right)
\right\} d^{4}x\mathbf{,}  \label{A.7}
\end{equation}
Here and in what follows, a summation over repeated Latin indices is
produced $(0-3)$.

Note that according to (\ref{A.4}), the relations (\ref{A.6}) can be written
in the form 
\begin{equation}
j^{k}=\left\{ \frac{\partial J}{\partial \xi _{0,0}},\frac{\partial J}{%
\partial \xi _{0,0}}\left( J^{-1}\frac{\partial J}{\partial \xi _{0,\alpha }}%
\right) \left( J^{-1}\frac{\partial J}{\partial \xi _{0,0}}\right)
^{-1}\right\} =\left\{ \rho ,\rho \frac{dx^{\alpha }}{dt}\right\} ,\qquad
\rho \equiv \frac{\partial J}{\partial \xi _{0,0}}  \label{A.7a}
\end{equation}
It is clear from (\ref{A.7a}) that $j^{k}$ is the 4-flux of particles, with $%
j^{0}=\rho $ being its density.

Variation of (\ref{A.7}) with respect to $\xi _{i}$ gives 
\begin{equation}
\frac{\delta \mathcal{A}_{\mathcal{E}_{\mathrm{st}}\left[ \mathcal{S}_{%
\mathrm{st}}\right] }}{\delta \xi _{i}}=-\partial _{l}\left( p_{k}\frac{%
\partial ^{2}J}{\partial \xi _{0,k}\partial \xi _{i,l}}\right) =-\frac{%
\partial ^{2}J}{\partial \xi _{0,k}\partial \xi _{i,l}}\partial
_{l}p_{k}=0,\qquad i=0,1,2,3  \label{A.8}
\end{equation}
Using identities 
\begin{equation}
\frac{\partial ^{2}J}{\partial \xi _{0,k}\partial \xi _{i,l}}\equiv
J^{-1}\left( \frac{\partial J}{\partial \xi _{0,k}}\frac{\partial J}{%
\partial \xi _{i,l}}-\frac{\partial J}{\partial \xi _{0,l}}\frac{\partial J}{%
\partial \xi _{i,k}}\right)   \label{A.9}
\end{equation}
\begin{equation}
\frac{\partial J}{\partial \xi _{i,l}}\xi _{k,l}\equiv J\delta
_{k}^{i},\qquad \partial _{l}\frac{\partial ^{2}J}{\partial \xi
_{0,k}\partial \xi _{i,l}}\equiv 0  \label{A.10}
\end{equation}
one can test by direct substitution that the general solution of linear
equations (\ref{A.8}) has the form 
\begin{equation}
p_{k}=\frac{b_{0}}{2}\left( \partial _{k}\varphi +g^{\alpha }\left( \mathbf{%
\xi }\right) \partial _{k}\xi _{\alpha }\right) ,\qquad k=0,1,2,3
\label{A.11}
\end{equation}
where $b_{0}\neq 0$ is an arbitrary constant, $g^{\alpha }\left( \mathbf{\xi 
}\right) ,\;\;\alpha =1,2,3$ are arbitrary functions of $\mathbf{\xi =}%
\left\{ \xi _{1},\xi _{2},\xi _{3}\right\} $, and $\varphi $ is the dynamic
variable $\xi _{0}$, which ceases to be fictitious. It is the conceptual
integration, which allows one to introduce the wave function. Let us
substitute (\ref{A.11}) in (\ref{A.7}). The term of the form $\partial
_{k}\varphi \partial J/\partial \xi _{0,k}$ is reduced to Jacobian and does
not contribute to dynamic equation. The terms of the form $\xi _{\alpha
,k}\partial J/\partial \xi _{0,k}$ vanish due to identities (\ref{A.10}). We
obtain 
\begin{equation}
\mathcal{A}_{\mathcal{E}_{\mathrm{st}}\left[ \mathcal{S}_{\mathrm{st}}\right]
}\left[ \varphi ,\mathbf{\xi },j\right] =\int \left\{ \frac{m}{2}\frac{%
j^{\alpha }j^{\alpha }}{j^{0}}-j^{k}p_{k}-\frac{\hbar ^{2}}{8m}\frac{\left( 
\mathbf{\nabla }\rho \right) ^{2}}{\rho }\right\} d^{4}x\mathbf{,\qquad }%
j^{0}=\rho   \label{A.12}
\end{equation}
where quantities $p_{k}$ are determined by the relations (\ref{A.11})

Variation of the action (\ref{A.12}) with respect to $j^{k}$ gives 
\begin{equation}
p_{0}=-\frac{m}{2}\frac{j^{\alpha }j^{\alpha }}{\rho ^{2}}+\frac{\hbar ^{2}}{%
8m}\left( \frac{\left( \mathbf{\nabla }\rho \right) ^{2}}{\rho ^{2}}+2%
\mathbf{\nabla }\frac{\left( \mathbf{\nabla }\rho \right) }{\rho }\right) 
\label{A.14}
\end{equation}
\begin{equation}
p_{\beta }=m\frac{j^{\beta }}{\rho },\qquad \beta =1,2,3  \label{A.17}
\end{equation}

Now we eliminate the variables $\mathbf{j}=\left\{ j^{1},j^{2},j^{3}\right\} 
$ from the action (\ref{A.12}), using relation (\ref{A.17}). We obtain 
\begin{equation}
\mathcal{A}_{\mathcal{E}_{\mathrm{st}}\left[ \mathcal{S}_{\mathrm{st}}\right]
}\left[ \rho ,\varphi ,\mathbf{\xi }\right] =\int \left\{ -p_{0}-\frac{%
p_{\beta }p_{\beta }}{2m}-\frac{\hbar ^{2}}{8m}\frac{\left( \mathbf{\nabla }%
\rho \right) ^{2}}{\rho ^{2}}\right\} \rho d^{4}x\mathbf{,}  \label{A.18}
\end{equation}
where $p_{k}$ is determined by the relation (\ref{A.11}).

Now instead of dependent variables $\rho ,\varphi ,\mathbf{\xi }$ we
introduce the $n$-component complex function $\psi $, defining it by
relations (\ref{s1.1}) -- (\ref{s5.5})

The function $\psi $ is constructed of the variable $\varphi $, the fluid
density $\rho $ and the Lagrangian coordinates $\mathbf{\xi }$, considered
as functions of $\left( t,\mathbf{x}\right) $, as follows \cite{R99}. The $n$%
-component complex function $\psi =\{\psi _{\alpha }\},\;\;\alpha
=1,2,\ldots ,n$ is defined by the relations 
\begin{equation}
\psi _{\alpha }=\sqrt{\rho }e^{i\varphi }u_{\alpha }(\mathbf{\xi }),\qquad
\psi _{\alpha }^{\ast }=\sqrt{\rho }e^{-i\varphi }u_{\alpha }^{\ast }(%
\mathbf{\xi }),\qquad \alpha =1,2,\ldots ,n,  \label{A.19}
\end{equation}
\begin{equation}
\psi ^{\ast }\psi \equiv \sum_{\alpha =1}^{n}\psi _{\alpha }^{\ast }\psi
_{\alpha },  \label{A.20}
\end{equation}
where (*) means the complex conjugate. The quantities $u_{\alpha }(\mathbf{%
\xi })$, $\alpha =1,2,\ldots ,n$ are functions of only variables $\mathbf{%
\xi }$, and satisfy the relations 
\begin{equation}
-{\frac{i}{2}}\sum_{\alpha =1}^{n}\left( u_{\alpha }^{\ast }\frac{\partial
u_{\alpha }}{\partial \xi _{\beta }}-\frac{\partial u_{\alpha }^{\ast }}{%
\partial \xi _{\beta }}u_{\alpha }\right) =g^{\beta }(\mathbf{\xi }),\qquad
\beta =1,2,3,\qquad \sum_{\alpha =1}^{n}u_{\alpha }^{\ast }u_{\alpha }=1.
\label{A.21}
\end{equation}
The number $n$ is such a natural number that equations (\ref{A.21}) admit a
solution. In general, $n$ depends on the form of the arbitrary integration
functions $\mathbf{g}=\{g^{\beta }(\mathbf{\xi })\}$, $\beta =1,2,3$. The
functions $\mathbf{g}$ determine vorticity of the fluid flow.

It is easy to verify that 
\begin{equation}
\rho =\psi ^{\ast }\psi ,\qquad \rho p_{0}\left( \varphi ,\mathbf{\xi }%
\right) =-\frac{ib_{0}}{2}(\psi ^{\ast }\partial _{0}\psi -\partial _{0}\psi
^{\ast }\cdot \psi )  \label{s5.6}
\end{equation}
\begin{equation}
\rho p_{\alpha }\left( \varphi ,\mathbf{\xi }\right) =-\frac{ib_{0}}{2}(\psi
^{\ast }\partial _{\alpha }\psi -\partial _{\alpha }\psi ^{\ast }\cdot \psi
),\qquad \alpha =1,2,3,  \label{s5.7}
\end{equation}
The variational problem with the action (\ref{A.18}) appears to be
equivalent to the variational problem with the action functional 
\[
\mathcal{A}_{\mathcal{E}_{\mathrm{st}}\left[ \mathcal{S}_{\mathrm{st}}\right]
}[\psi ,\psi ^{\ast }]=\int \left\{ \frac{ib_{0}}{2}(\psi ^{\ast }\partial
_{0}\psi -\partial _{0}\psi ^{\ast }\cdot \psi )\right. 
\]
\begin{equation}
+\left. \frac{b_{0}^{2}}{8m\rho }(\psi ^{\ast }\mathbf{\nabla }\psi -\mathbf{%
\nabla }\psi ^{\ast }\cdot \psi )^{2}-\frac{\hbar ^{2}}{8m}\frac{\left( 
\mathbf{\nabla }\rho \right) ^{2}}{\rho }\right\} \mathrm{d}^{4}x.
\label{s5.8}
\end{equation}

We hope that in the case $n=2$ equations (\ref{A.21}) have a solution for
any functions $\mathbf{g}$, because in this case the number (four) of real
components of $\psi $ coincides with the number of hydrodynamic variables $%
j^{k}$ ($k=0,,2,3)$. (But this statement is not yet proved). For the
two-component function $\psi $ ($n=2$) the following identity takes place 
\begin{equation}
(\mathbf{\nabla }\rho )^{2}-(\psi ^{\ast }\mathbf{\nabla }\psi -\mathbf{%
\nabla }\psi ^{\ast }\cdot \psi )^{2}\equiv 4\rho \mathbf{\nabla }\psi
^{\ast }\mathbf{\nabla }\psi -\rho ^{2}\sum\limits_{\alpha =1}^{\alpha
=3}\left( \mathbf{\nabla }s_{\alpha }\right) ^{2},  \label{s5.30}
\end{equation}
\begin{equation}
\rho \equiv \psi ^{\ast }\psi ,\qquad s\equiv \frac{\psi ^{\ast }\mathbf{%
\sigma }\psi }{\rho },\qquad \mathbf{\sigma }=\{\sigma _{\alpha }\},\qquad
\alpha =1,2,3,  \label{s5.31}
\end{equation}
where $\sigma _{\alpha }$ are the Pauli matrices. In virtue of the identity (%
\ref{s5.30}) the action (\ref{s5.8}) reduces to the form 
\begin{eqnarray}
\mathcal{E}_{\mathrm{st}}\left[ \mathcal{S}_{\mathrm{st}}\right] &:&\qquad 
\mathcal{A}_{\mathcal{E}_{\mathrm{st}}\left[ \mathcal{S}_{\mathrm{st}}\right]
}\left[ \psi ,\psi ^{\ast }\right] =\int \left\{ \frac{ib_{0}}{2}\left( \psi
^{\ast }\partial _{0}\psi -\partial _{0}\psi ^{\ast }\cdot \psi \right) -%
\frac{b_{0}^{2}}{2m}\mathbf{\nabla }\psi ^{\ast }\mathbf{\nabla }\psi \right.
\nonumber \\
&&\left. +\frac{b_{0}^{2}}{8m}\sum\limits_{\alpha =1}^{\alpha =3}(\mathbf{%
\nabla }s_{\alpha })^{2}\rho +\frac{b_{0}^{2}-\hbar ^{2}}{8\rho m}(\mathbf{%
\nabla }\rho )^{2}\right\} d^{4}x,  \label{s5.32}
\end{eqnarray}
where $\mathbf{s}$ and $\rho $ are defined by the relations (\ref{s5.31}).
Thus, we prove the relation (\ref{c4.18}).

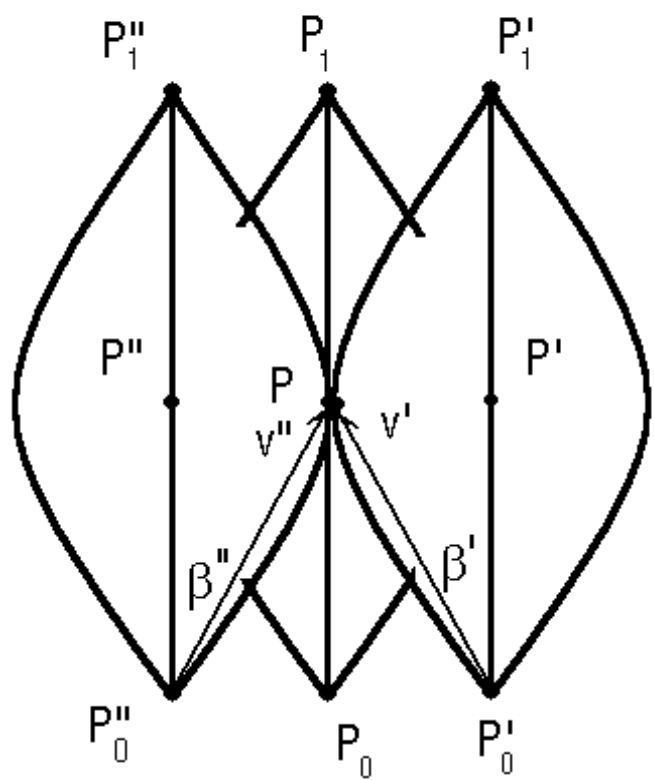
\end{document}